\documentclass[a4paper,11pt]{article}
\usepackage{jheppub} %
\usepackage{lineno, subfigure}
\usepackage[ruled,vlined]{algorithm2e}
\usepackage{algpseudocode}
\usepackage{caption}
\usepackage[nice]{nicefrac}
\usepackage[normalem]{ulem}
\usepackage{xcolor,colortbl}
	
\usepackage{color, colortbl}
\definecolor{LightCyan}{rgb}{1,0.88,1}
\definecolor{LightGrey}{rgb}{0.85,0.85,0.85}

\newcommand{\mO}{\mathcal{O}}

\title{\boldmath Numerical tests of the large charge expansion}

\author[a,b]{Gabriel Cuomo,}    
\author[c,d,e]{J. M. Viana Parente Lopes,}
\author[c,d,e,f]{José Matos,}
\author[c,d,e]{Júlio Oliveira,}
\author[f]{and João Penedones}

\affiliation[a]{Simons Center for Geometry and Physics, SUNY, Stony Brook, NY 11794, USA}      
\affiliation[b]{C. N. Yang Institute for Theoretical Physics, Stony Brook University, Stony Brook, NY 11794, USA}  
\affiliation[c]{Associate Laboratory LaPMET, 4169-007 Porto, Portugal.}
\affiliation[d]{Departamento de Física e Astronomia, Faculdade de Ciências, Universidade do Porto, rua do Campo Alegre s/n, 4169-007 Porto, Portugal.}
\affiliation[e]{Centro de Física das Universidades do Minho e Porto (CF-UM-PT), Departamento de Física e Astronomia, Faculdade de Ciências, Universidade do Porto, 4169-007 Porto, Portugal.}
\affiliation[f]{Fields and Strings Laboratory, Institute of Physics, École Polytechnique Fédéral de Lausanne (EPFL), Route de la Sorge, CH-1015 Lausanne, Switzerland}

\emailAdd{jose.bouradematos@gmail.com}

\abstract{We perform Monte-Carlo measurements of two and three-point functions of charged operators in the critical $O(2)$ model in 3 dimensions. 
Our results are compatible with the predictions of the \emph{large charge} superfluid effective field theory. To obtain reliable measurements for large values of the charge, we improved the Worm algorithm and devised a measurement scheme which mitigates the uncertainties due to lattice and finite size effects.
}

\begin{document}
\maketitle
\flushbottom

\section{Introduction}\label{sec:intro}

Conformal field theories (CFTs) play a key role in physics. CFTs are fixed points of the Renormalization Group flow and encode the universal properties of critical points in second-order phase transitions. Additionally, CFTs shed light on some of the mysteries of quantum gravity through the AdS/CFT correspondence. %

Local observables in CFTs can be described in terms of a set of dimensionless numbers (the CFT data), i.e. the conformal dimensions and OPE coefficients of the primary fields of the theory \cite{Rychkov:2016iqz,Simmons-Duffin:2016gjk}. When the CFT is strongly coupled, the spectrum of low-dimension operators often  lacks an  organizational principle; consequently one has to resort to numerical methods, such as the bootstrap %
or Monte Carlo simulations, to compute the corresponding CFT data.

It has been recently realized that sectors with large quantum numbers are often amenable to a perturbative description. 
Notably, CFTs become weakly coupled in the large spin sector \cite{Alday:2007mf}, in the sense that the CFT data can be computed in an expansion in inverse powers of the spin via the analytic bootstrap \cite{Fitzpatrick:2012yx,Komargodski:2012ek}. Relevant to our work is the large charge expansion \cite{Hellerman:2015nra,Monin:2016jmo}, which applies to operators charged under the internal symmetries of the theory. Both of these expansions, besides being useful per se, allow identifying interesting patterns in the CFT spectrum, with operators naturally organized in \emph{Regge trajectories} as a function of their quantum numbers.\footnote{Remarkably, it has been recently proven that the CFT data admit rigorous analyticity properties as a function of the spin of the corresponding operators \cite{Caron-Huot:2017vep,Simmons-Duffin:2017nub}.}

Let us review the physical picture underlying the large charge expansion.
Consider for concreteness a CFT invariant under an internal $U(1)$ symmetry group in three spacetime dimensions.
By the state-operator correspondence, an operator with a large $U(1)$ charge is associated with a finite density state for the theory quantized on the cylinder. In \cite{Hellerman:2015nra} it was argued that, in the simplest case, the corresponding state is found in a superfluid phase. 
In this case, one can describe large charge states via the \emph{universal} effective field theory (EFT) description for the \emph{hydrodynamic} Goldstone mode of the superfluid \cite{Monin:2016jmo}.\footnote{Here we are focusing on \emph{generic} theories, where no additional symmetries are present and thus all other (\emph{radial}) modes are gapped at finite density; a well-studied exception is given by supersymmetric CFTs with moduli spaces \cite{Hellerman:2017veg,Hellerman:2018xpi,Grassi:2019txd,Sharon:2020mjs}, where there are additional light modes in the spectrum.} This allows for the systematic calculation of correlation functions of charged operators, where the derivative expansion coincides with an expansion in inverse powers of the charge.

The superfluid EFT is believed to describe the large charge sector of a vast class of theories. Nonetheless, sometimes other phases are possible, e.g. Fermi spheres in fermionic theories \cite{Komargodski:2021zzy,Dondi:2022zna} or extremal Reissner-Nordstorm black holes in holographic models. Furthermore, in contrast with the large spin expansion, even in specific theories there is no rigorous bootstrap proof of the validity of the superfluid description.\footnote{See however \cite{Jafferis:2017zna} for interesting progress in this direction.} It is therefore important to check its predictions in theories where explicit calculations are possible. The main purpose of this paper is to provide evidence for the validity of the superfluid EFT in a specific strongly coupled CFT, namely the $O(2)$ model in three dimensions, via Monte-Carlo calculations.\footnote{Notice that checking the superfluid EFT provides also an indirect test for the validity of the state-operator correspondence in the $O(2)$ model. Recently, the state-operator map was directly tested numerically for the Ising model in \cite{Zhu:2022gjc}.}

\subsection{Background and summary}

Let us begin discussing some of the predictions of the large charge expansion, with a particular focus on the features which are specific to the superfluid EFT. 
The scaling dimension $D(Q)$ of the operator $\mO_Q$ with lowest dimension at fixed charge $Q$ is given by \cite{Hellerman:2015nra,Monin:2016jmo}
\begin{equation}\label{eq:conformal dimension of charged scalars}
D\left(Q\right) =c_{{\nicefrac{3}{2}}}Q^{3/2}+c_{\frac{1}{2}}Q^{1/2}+c_{0}+O\left(Q^{-1/2}\right).
\end{equation}
The coefficients $c_{3/2}$ and $c_{1/2}$ in \eqref{eq:conformal dimension of charged scalars} are model-dependent Wilson coefficients, while $c_0\simeq-0.0937$. The leading behaviour at large $Q$ follows from dimensional analysis. The existence of an expansion in $1/Q$ is less trivial, but it is not specific to the superfluid EFT only; for instance, the result for a Reissner-Nordstrom black hole admits a similar expansion (see e.g. \cite{Loukas:2018zjh}). The $Q^0$ contribution $c_0\simeq-0.0937$ is associated with the Casimir energy of the Goldstone and it is thus a specific prediction of the superfluid phase \cite{Monin:2016bwf}. 

The superfluid EFT also predicts other observables in terms of the \textbf{same} Wilson coefficients. For instance, it predicts that the primary operator with the next-to-lowest dimension with charge $Q$ has spin $2$ and scaling dimension $D(Q)+\sqrt{3}$ \cite{Hellerman:2015nra}. Importantly for this work, the EFT also predicts OPE coefficients of the operators $\mO_Q$ \cite{Monin:2016jmo,Cuomo:2020rgt,Cuomo:2021ygt,Dondi:2022wli}, defined as 
\begin{equation}\label{eq_OPE_def}
\lambda_{Q_{1},Q_{2},-Q_{1}-Q_{2}}=\lim_{|x| \to \infty} |x|^{D(Q_1+Q_2)}\langle \mO_{Q_{1}}(0)\mO_{Q_{2}}(1)\mO_{-Q_{1}-Q_{2}}(x)\rangle\,.
\end{equation}
The structure of the prediction depends on whether one considers three large charge operators or two large charge operators and one with a small charge. We refer to these predictions, respectively, as OPE in Regime I and Regime II. The results are
\begin{itemize}
\item \textbf{Regime I:} the prediction for the OPE of three large charge operators reads \cite{Cuomo:2021ygt}
\begin{equation}
\lambda_{Q_{1},Q_{2},-Q_{1}-Q_{2}}=\exp\left[\frac{c_{{\nicefrac{3}{2}}}}{2\sqrt{\pi}}Q_{1}^{3/2}f\left(y\right)+O\left(\min\left\{ Q_{1},Q_{2}\right\} ^{1/2}\right)\right],\quad y=\sqrt{Q_{2}/Q_{1}},\label{eq:regime I asymptotic behavior}
\end{equation}
where $Q_{1},Q_{2}\gg1$, and $f\left(y\right)$ is given in terms of the solution of a non-linear PDE. We shall only need its value at $y=1$, $f\left(1\right)\simeq 0.996$. Notice that this prediction depends only on $c_{{\nicefrac{3}{2}}}$ to leading order in the derivative expansion.
\item \textbf{Regime II:} the prediction for the OPE of two large charge operators and a small charge operator is \cite{Monin:2016jmo,Cuomo:2020rgt,Dondi:2022wli}
\begin{equation}\label{eq:regime II asymptotic behavior}
\lambda_{Q_{1},Q_{2},-Q_{1}-Q_{2}}=C\left(Q_{2}\right)Q_{1}^{D(Q_2)/2}\left[1+0.46c_{{\nicefrac{3}{2}}}\times\frac{Q_{2}^{2}}{\sqrt{Q_{1}}}+O\left(1/Q_{1}\right)\right].
\end{equation}
where $Q_{1}\gg1$ and $Q_{2}=O(1)$. The coefficient $C(Q_2)$ is a novel Wilson coefficient, associated with the operator matching for $\mO_{Q_2}$ in terms of the superfluid Goldstone \cite{Monin:2016jmo}. As in \eqref{eq:conformal dimension of charged scalars}, the leading scaling with $Q_1$ can be inferred from dimensional analysis \cite{Jafferis:2017zna}. The subleading correction depends on $c_{{\nicefrac{3}{2}}}$ and it is a specific prediction of the superfluid EFT. The numerical coefficient multiplying $c_{{\nicefrac{3}{2}}}$ does not admit a simple analytic expression. It was computed in \cite{Cuomo:2020rgt} from the (small) shift of the superfluid saddle-point (equivalent to a tadpole diagram) due to the charge $Q_2$ sourced by the operator insertion.\footnote{The result for this OPE reported in \cite{Dondi:2022wli} differs by a factor of $2$ because of a typo; we thank Nicola Dondi for checking the result.}  
\end{itemize}

We now discuss former tests of the large charge expansion in CFTs. 
The validity of the EFT has been unambiguously demonstrated in several perturbative theories, see e.g. \cite{DeLaFuente:2018uee,Badel:2019khk,Sharon:2020mjs,Antipin:2022naw};
particularly relevant for us are the results for large charge operators in the $O(N)$ model in the $\varepsilon$-expansion \cite{Badel:2019oxl,Antipin:2020abu} and at large $N$ \cite{Alvarez-Gaume:2019biu,Giombi:2020enj} 

It is of course harder to study directly strongly coupled CFTs.   \cite{Banerjee:2017fcx} initiated the Monte-Carlo study of the large charge sector of the $O(2)$ model computing the scaling dimension $D(Q)$ for $Q=1,2,\ldots, 12$. To perform the calculations the authors applied the worm algorithm \cite{Prokofev:2001ddj} to the worldline formulation of the classical $O(2)$ sigma-model \cite{Banerjee:2010kc} (see appendix \ref{sec:Monte Carlo} for details). Remarkably, the numerical results agree with the theoretical prediction \eqref{eq:conformal dimension of charged scalars} up to $Q=O(1)$, providing a determination of the coefficients $c_{{\nicefrac{3}{2}}}$ and $c_{\frac{1}{2}}$ from the fit (with $c_0$ assumed as input). Similar calculations have been performed in the $O(N)$ models for $N=4$ \cite{Banerjee:2019jpw,Banerjee:2021bbw} and $N=6,8$ \cite{Singh:2022akp}.

The results of \cite{Banerjee:2017fcx} provide strong evidence for the existence of a $1/Q$ expansion for $D(Q)$. This is a very nontrivial result, but, as commented earlier, it is not necessarily specific to the superfluid EFT (even if it is admittedly hard to think of alternative descriptions for the large charge sector of the $O(2)$ model). The main goal of this work is to test specifically the superfluid EFT by studying OPE coefficients of charged operators. Below we give a brief summary of our results.

First, in sec.~\ref{sec:conformal dimension of scalar operators} we compute the scaling dimension $D(Q)$ for charges up to $Q=19$, thus extending the pre-existing results for $Q\leq 12$ \cite{Banerjee:2017fcx}. The results are plotted in fig.~\ref{fig:conformal dimension}. Our measurements are compatible with those of \cite{Banerjee:2017fcx} and provide improved estimates for the values of the Wilson coefficients $c_{{\nicefrac{3}{2}}}$ and $c_{\frac{1}{2}}$ in the $O(2)$ model, cfr. eq.~\eqref{eq:results conformal dimension}. Unfortunately, we are not able to reach the precision needed to obtain a reliable estimate for the coefficient $c_0$ in eq.~\eqref{eq:conformal dimension of charged scalars}; our results are nonetheless compatible with the theoretically predicted value.

In order to test the superfluid EFT, in sec~\ref{sec:OPE coefficients} we study the OPE coefficients in eq.~\eqref{eq:regime I asymptotic behavior} and \eqref{eq:regime II asymptotic behavior}. Notice that extracting three-point functions from Monte-Carlo simulations is significantly more involved than computing two-point functions. In Regime I we computed the OPE coefficient for $Q_1=Q_2=1,2,3,4$, while in Regime II we obtained results for $Q_1=1,2,3,4,5$ with $Q_2=1$ (fixed). The results are shown in fig.~\ref{fig: OPE raw}. Despite the relative smallness of the charges we find good agreement between the numerical results and the EFT predictions, in both regimes. In particular, from the extrapolation of the result in Regime I to larger values of the charges we extract the coefficient $c_{{\nicefrac{3}{2}}}$, finding remarkable agreement with the value extracted from the measurement of the scaling dimension. The comparison is shown in fig.~\ref{fig:log OPE}. From the results in Regime II we measure the value of the coefficient $C(1)$ in eq.~\eqref{eq:regime II asymptotic behavior}, see fig.~\ref{fig:leading behavior regime II}. The estimate for $c_{{\nicefrac{3}{2}}}$ extracted from Regime II is encouragingly compatible with the one obtained from $D(Q)$ and the OPE coefficient in Regime I, but uncertainties are too large for our analysis to be conclusive; see fig.~\ref{fig:lambda 1 and lambda2}.

Overall our results provide encouraging evidence for the validity of the superfluid EFT in the large charge sector of the $O(2)$ model, but additional data would be helpful to unambiguously confirm the EFT description.
In sec.~\ref{sec:conclusions} we further speculate on the implications of our findings and comment on possible future directions.

To compute the correlation functions numerically we used the worm algorithm. We introduced two technical improvements with respect to the strategy of \cite{Banerjee:2017fcx}. First, we introduced the continuous time update step, which reduces the computational time; details are given in appendix \ref{sec:Monte Carlo}. Additionally, we devised an improved procedure to take the continuum limit. To this aim, we carefully analysed lattice effects, combining numerical experiments and conformal perturbation theory; some details are given in appendix \ref{sec:lattice corrections}.

\section{Conformal dimension of lightest charged scalar operator}\label{sec:conformal dimension of scalar operators}

Measuring 2pt functions of operators with large conformal dimensions is challenging. On the one hand, the 2pt function decays quickly when the distance between the operators increases.
On the other hand, measurements at short distances are contaminated by large lattice effects.

\begin{figure}[t]
\centering
\includegraphics[width=0.75\columnwidth]{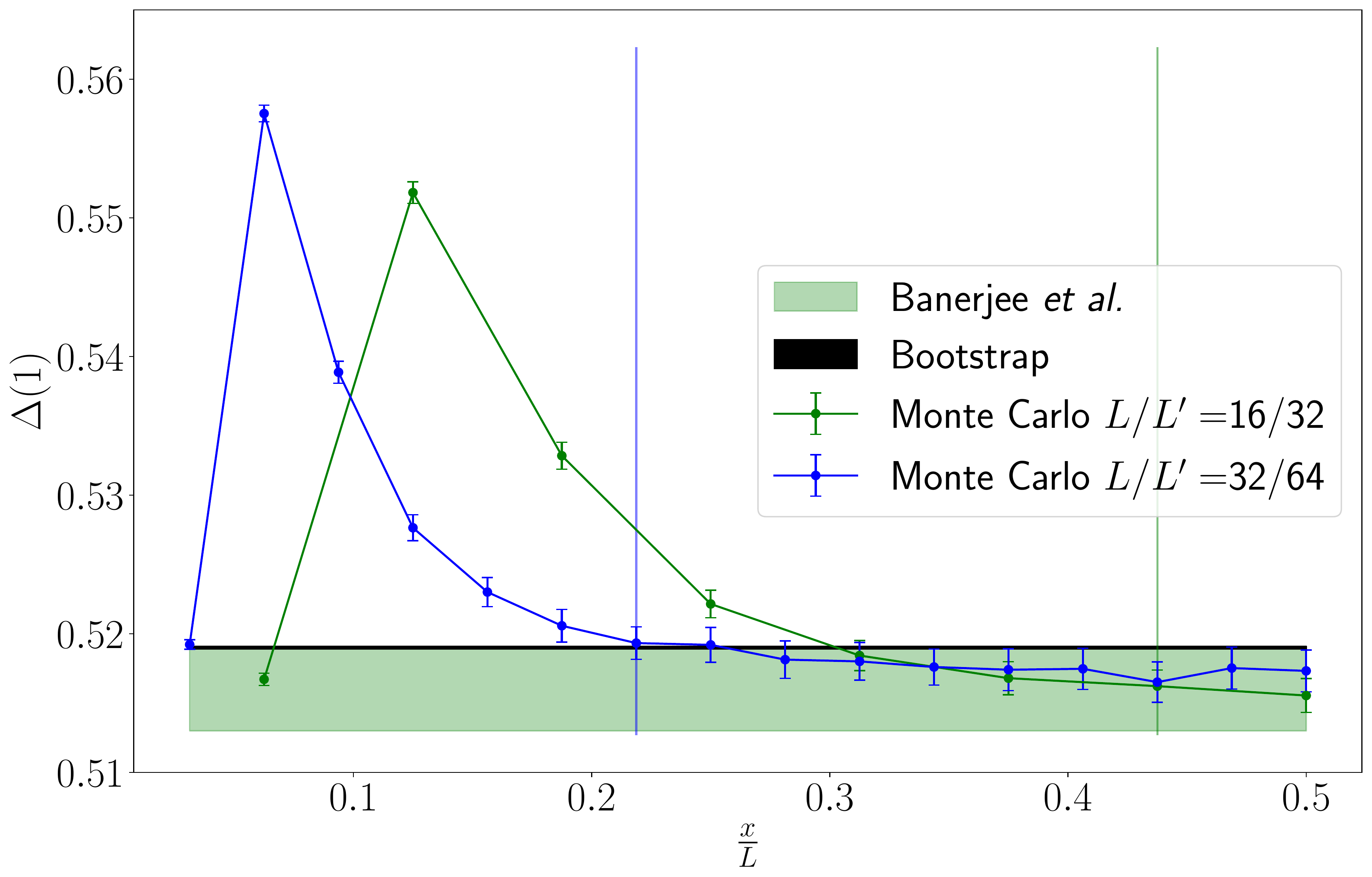}
\caption{$\Delta(1)$, extracted with eq.\eqref{eq:ratios}, for  $\alpha =2$ and $L\in[16,32]$. Notice that the region where there are significant deviations from a constant value, on the left of the vertical line, gets smaller as $L$ increases. This is 
expected from lattice effects. We include the previous Monte-Carlo result with error bars at 1$\sigma$, taken from Tab.I in \cite{Banerjee:2017fcx}, as well as the bootstrap result \cite{Kos:2015mba}. 
\label{fig:Ratios correlation function in the torus}}
\end{figure}

To make progress,   \cite{Banerjee:2017fcx} introduced a method that does not require sampling directly 2pt functions.
They measure the difference between the conformal dimensions of operators with consecutive charges, $\Delta(Q)\equiv D(Q+1) - D(Q)$, which scales as $Q^{1/2}$ instead of $Q^{3/2}$.
This is achieved by rewriting the 2pt function $C_{Q}\left(x\right)\equiv\left\langle \mO_Q(0)\mO_{-Q}(x)\right\rangle $ as a product of ratios
\begin{equation}\label{eq:decomposition 2pt function in ratios}
    C_{Q}\left(x\right)=\prod_{q=1}^{Q}R_{q}\left(x\right)\,, \qquad R_{q}\left(x\right)\equiv\dfrac{C_{q}\left(x\right)}{C_{q-1}\left(x\right)}\sim \dfrac{1}{|x|^{2\Delta(q)}}\,,\qquad C_0(x) \equiv 1.
\end{equation}
This is useful because, due to the worldline formulation of the $O(2)$ model, $R_q(x)$, can be sampled directly by computing the expectation value of operators with $Q=1$ in a background charge distribution
\begin{equation}\label{eq_Rq_MC}
R_{q}\left(x\right)=\left\langle e^{i\theta(0)}e^{-i\theta(x)}\right\rangle _{\left(q-1\right)_{0}-\left(q-1\right)_{x}},
\end{equation}
where $e^{i \theta}$ ($e^{-i\theta}$) represents a charge $1$ ($-1$) operator in the nonlinear sigma model and the subscript indicates that the expectation value is computed in the presence of charge $(q-1)$ at the origin $0$ and charge $-(q-1)$ at position $x$. Check App.~\ref{sec:ratio between correlation functions.} for details. 

Having reconstructed the 2pt function, the next step is to extract $\Delta(Q)$. The naive approach is to compute $R_q(x)$ for different values of $x$, take the log and fit the slope. Unfortunately, this cannot be done systematically due to both finite-size effects and lattice effects. %
 Lattice effects are due to the discrete nature of the lattice. This introduces another distance scale, the lattice spacing $a$, such that in the region where $x/a\sim O(1)$, the discrete nature of the lattice spoils the CFT predictions. We set $a=1$ in the following unless specified otherwise.

In summary, there is an intermediate region, where $x/a \gg 1$ and $x/L \ll 1/2$, such that the continuum infinite size CFT predictions hold. As we show next, we are able to drop the second restriction through a choice of observable that eliminates finite-size effects. 

\begin{figure}[t]
\centering
\begin{subfigure}
    \centering
    \includegraphics[width=0.75\columnwidth]{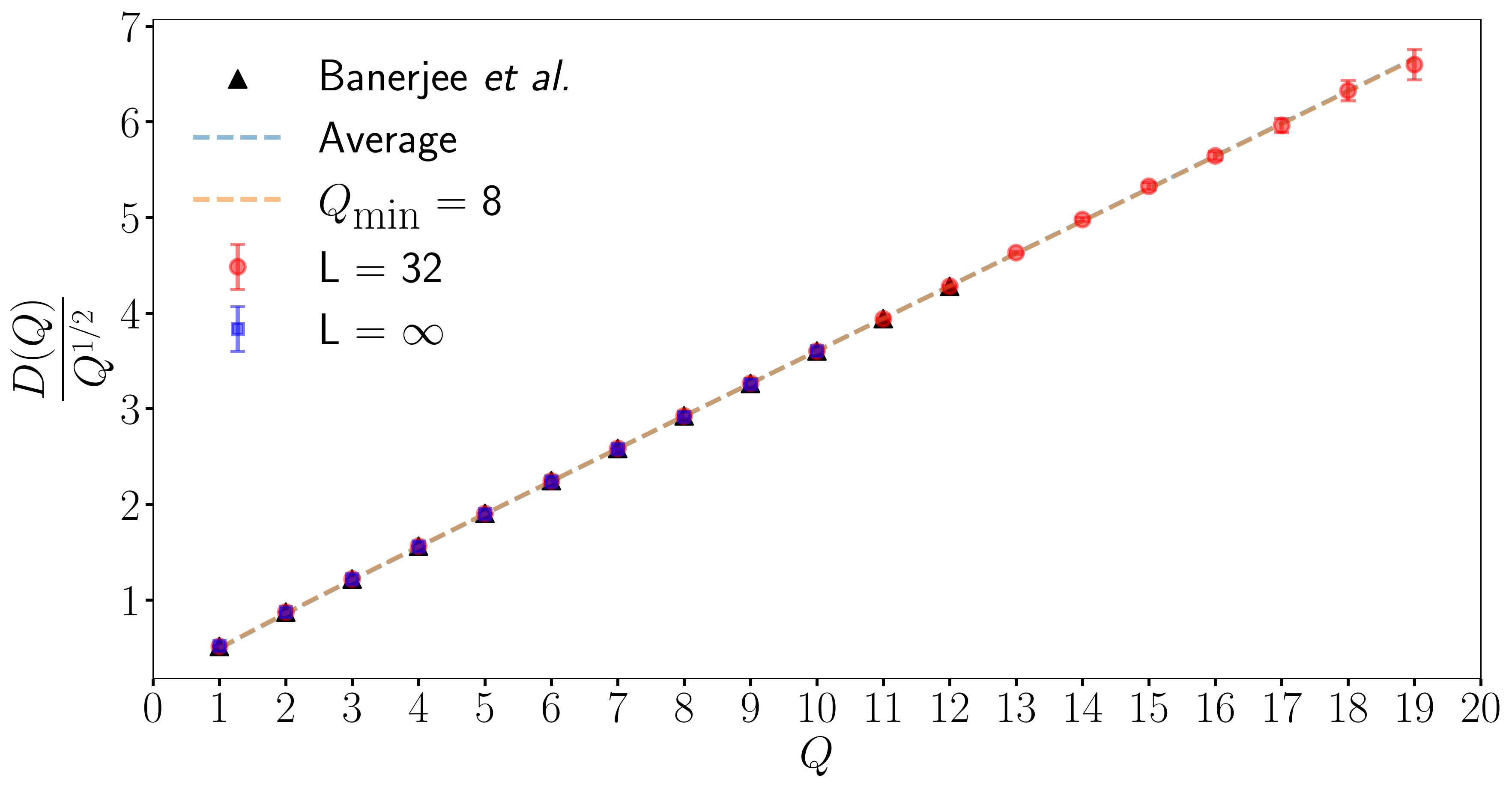}
    \caption{Here we plot both the measurements of $D(Q)/Q^{1/2}$ obtained for $L=32$ ($10^7$ Worm steps), as well as the results of the extrapolation to $L=\infty$, obtained using $\alpha=2$ in eq.~\eqref{eq:ratios}. We test the leading behaviour of eq. \eqref{eq:conformal dimension of charged scalars} by demonstrating the linearity of $D(Q)/Q^{1/2}$. We reproduce the previous results of \cite{Banerjee:2017fcx}. As explained at the end of the section, we performed best-fits of the coefficients $c_{{\nicefrac{3}{2}}}$ and $c_{\frac12}$ in eq.~\eqref{eq:conformal dimension of charged scalars}, restricting to charges $Q\geq Q_{min}$ for different choices of $Q_{min}$. The blue dashed curve is obtained using the coefficients extracted from the average of all the fits with $Q_{min}=4,\ldots,10$. The orange dashed curve is the best fit obtained using only charges $Q\geq 8$. Remarkably, the two lines are almost indistinguishable and almost overlap with all the data points.
    In the table below we list the
    data used in the fits presented in this section. Check app. \ref{sec:extrapolation} for a detailed discussion. }
    \label{fig:conformal dimension}
\end{subfigure}
\vspace*{0.2em}
\begin{subtable}
    \centering
    \begin{tabular}{|c|c|c|c|c|c|c|c|}
    \hline 
    \rowcolor{LightGrey} $Q$ & 1 & 2 & 3 & 4 & 5 & 6 & 7  \tabularnewline
    \hline 
    $D(Q)$ & 0.5201(9) & 1.236(1) & 2.111(2) & 3.120(4) & 4.250(6) & 5.484(9) & 6.82(1)  \tabularnewline
    \hline \hline
    \rowcolor{LightGrey}  $Q$ & 8 & 9 & 10 & 11 & 12 & 13 & 14   \tabularnewline
    \hline 
    $D(Q)$ & 8.28(2) & 9.80(2) & 11.39(3) & 13.07(4) & 14.82(6) & 16.7(1) & 18.6(2)  \tabularnewline
    \hline \hline
    \rowcolor{LightGrey}  $Q$& 15 & 16 & 17 & 18 & 19 &  &   \tabularnewline
    \hline 
    $D(Q)$ & 20.6(3) & 22.6(3) & 24.6(6) & 26.8(9) & 29(1) &  &   \tabularnewline
    \hline
    \end{tabular}
    \end{subtable}
    \vspace*{0em}
\end{figure}

Systematic errors coming from finite size effects are eliminated by computing the ratio between 2pt functions measured for different lattice sizes but at the same relative position
\begin{equation}\label{eq:ratios}
\dfrac{C_{Q,L}(x)}{C_{Q,\alpha L}(\alpha x)} = \alpha^{2D(Q)}\quad \text{or} \quad \dfrac{R_{Q,L}(x)}{R_{Q,\alpha L}(\alpha x)} = \alpha^{2\Delta(Q)}.
\end{equation}
The right-hand side holds as long as lattice effects are negligible. Notice that these relations are independent of the position $x$. Both ratios are insensitive to finite size effects since these are parameterized by the relative position $x/L$. Deviations from a constant value (as a function of $x$) are a proxy for lattice effects.

In fig.~\ref{fig:Ratios correlation function in the torus} %
we present the measurements of $\Delta(1)$, obtained through eq.~\eqref{eq:ratios}. For small values of $x$ there are deviations as expected. These are the lattice effects. They disappear when $x \approx 7$, identified by the vertical lines. In particular, we can also reliably measure the correlation function for distances $x < L/2$, which we will use to optimize our measurements. %
In the constant region, to the right of the vertical lines, we have an independent estimate of $\Delta(1)$ for each position. From these, we estimate the error bars on the final result.

The generalization of this analysis to different values of $Q$ provides reliable measurements of $\Delta(Q)$, which are not susceptible to finite-size effects and do not require fits. Similarly, we also obtain accurate estimates of the errors introduced by lattice effects. The measurements of $\Delta(Q)$ are independent of the lattice size. We explicitly checked this for larger charges. The results for $L=32$ and $L=64$ match within the statistical uncertainties, but $L=32$ has a smaller error\footnote{This is because the computational time required to perform a certain number of \emph{worm steps} increases with the lattice size; a smaller lattice size, therefore, allows to obtain measurements with higher statistical significance.}
and we focused on this system size.

We measured $D(Q)$ up to $Q=19$, see fig.~\ref{fig:conformal dimension}. As remarked in the introduction, this represents a considerable improvement with respect to the existing results, which stopped at $Q=12$ \cite{Banerjee:2017fcx}. To obtain results for such high values of the charge we used a continuous-time update step (see App.~\ref{sec:Monte Carlo}). Additionally, we sampled the ratio in eq.~\eqref{eq:ratios} for relatively small values of the distance $x$\footnote{We control for lattice effects by checking the dependence of the different estimates of $D(Q)$ on the position.}, while   \cite{Banerjee:2017fcx} performed all measurements for $x\sim L/2$.
Indeed, as explained earlier, lattice effects are negligible already for $x\gtrsim 7$, with the most precise measurements obtained for $7 \lesssim x \ll L/2$.\footnote{A coarse estimate for the precision of a measurement is $1/\sqrt{N}$, where $N$ is the number of Worm steps. Then the relative error of $R_q(x)$ should be of order $(1/\sqrt{N}) / R_q(x) \sim  x^{2\Delta(Q)} /\sqrt{N}$. Thus, for larger $\Delta(Q)$, it is important that we can restrict to small $x$, since $N$ is limited by the available computational resources.}

Let us now discuss the comparison with the theoretical prediction eq.~\eqref{eq:conformal dimension of charged scalars}. In doing so, we face some important questions: What is the theoretical error of the large charge expansion? Is this error also under control for small values of Q? 

\begin{figure}
    \centering
    \includegraphics[width = 0.5\columnwidth]{"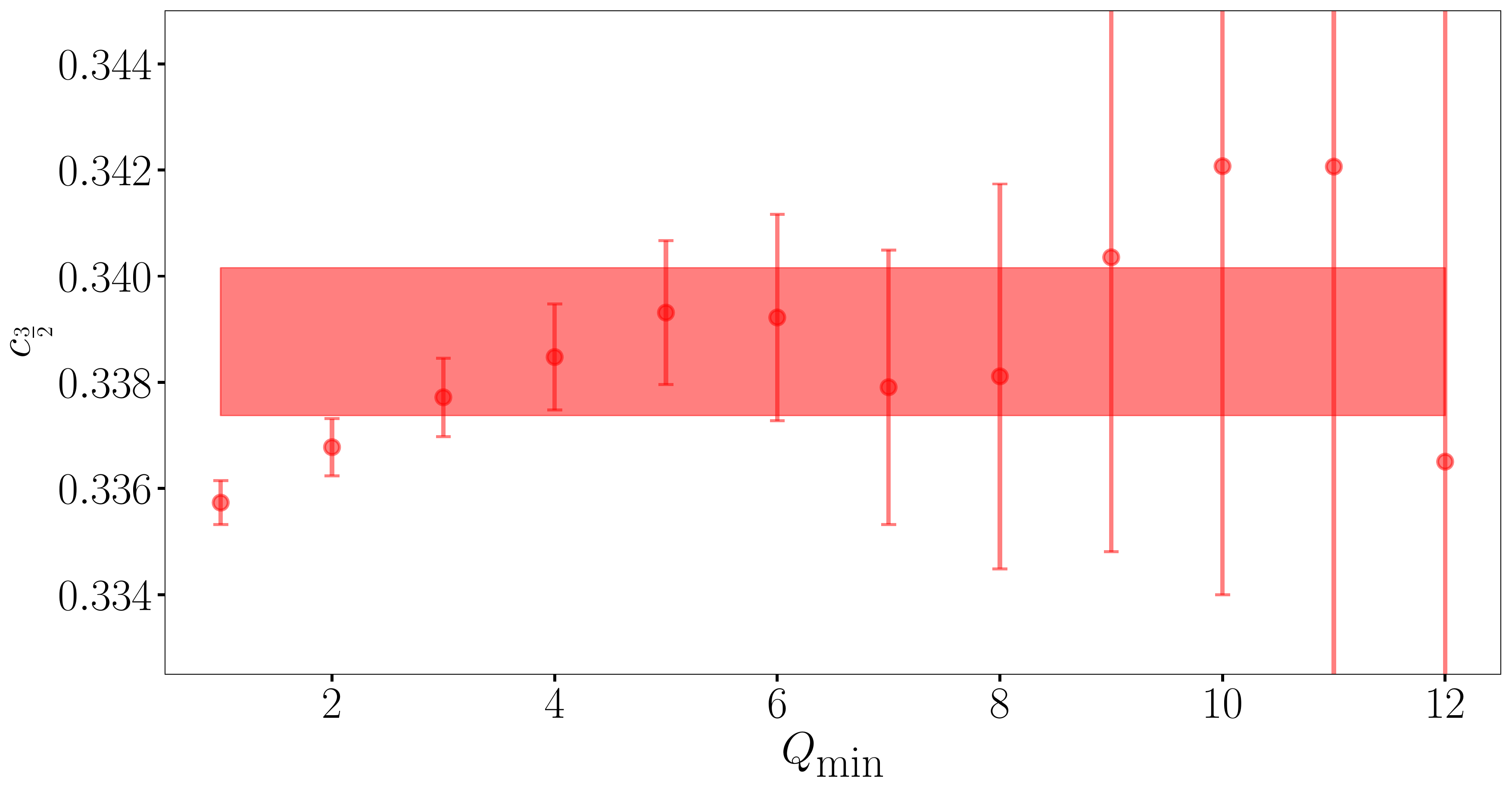"}\includegraphics[width = 0.5\columnwidth]{"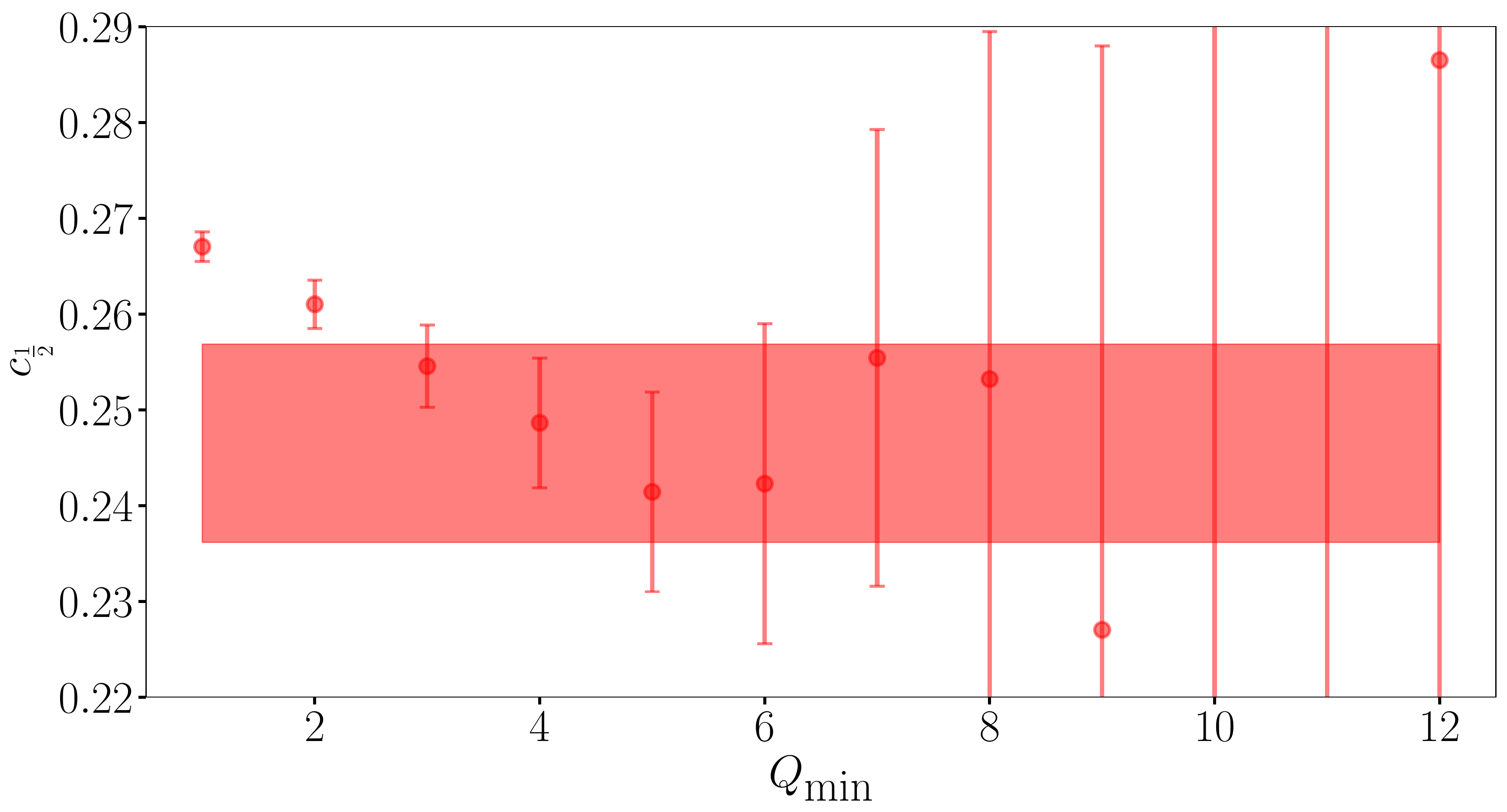"}
    \caption{Best-fit values of $c_{\nicefrac{3}{2}}$ (left) and $c_{\nicefrac{1}{2}}$ (right) as a function of the minimum charge included in the fit. For larger values of $Q_\textrm{min}$ the error bars are larger than the plotted range. The coloured regions represent the 1$\sigma$ interval quoted on \eqref{eq:results conformal dimension}. }
    \label{fig:q_min}
\end{figure}

The large charge expansion is believed to be an asymptotic expansion \cite{Dondi:2021buw}. The series in eq.~\eqref{eq:conformal dimension of charged scalars} thus includes both \emph{perturbative} terms, suppressed by inverse powers of $Q$,\footnote{Sometimes these power corrections may be enhanced by logarithms of the charge, see \cite{Cuomo:2020rgt}.} as well as \emph{non-perturbative} corrections, which are exponentially suppressed at large charge $\sim e^{-\# \sqrt{Q}}$.\footnote{See \cite{Grassi:2019txd,Hellerman:2021duh} for some progress in understanding similar corrections in supersymmetric theories.} Most importantly, as typical with asymptotic expansions, the series is not expected to converge to the exact result upon including infinitely many terms; rather, the large $N$ analysis of   \cite{Dondi:2021buw} suggests that the large charge expansion of the scaling dimension $D(Q)$ admits an optimal truncation after $n\sim \sqrt{Q}$ terms.

While the subtleties associated with the asymptotic nature of the series are unimportant for very large charges, they make it challenging to estimate the accuracy of the expansion for our data. In particular, we do not expect the theoretical error to be a simple function of $Q$ when $Q\sim O(1)$.

To (partially) account for these effects, we performed fits of $c_{\nicefrac{3}{2}}$ and $c_{\nicefrac{1}{2}}$ in eq.~\eqref{eq:conformal dimension of charged scalars} using charges in the range $[Q_{\text{min}}, 19]$ for different values of $Q_{\text{min}}$. We indeed expect that the theoretical error decreases with $Q$. The results are shown in fig.~\ref{fig:q_min}. The results are independent of $Q_\text{min}$ for $Q_\text{min} \gtrsim 3$, suggesting that the truncated asymptotic expansion is trustworthy beyond this value of the charge. By averaging over the results for $Q_\text{min}\in[4,8]$ we obtain 
\begin{equation}\label{eq:results conformal dimension}
    c_{\nicefrac{3}{2}} = 0.339(1) \qquad c_{\nicefrac{1}{2}} = 0.25(1).
\end{equation}
These values are compatible with the previous estimates of \cite{Banerjee:2017fcx} $c_{\nicefrac{3}{2}}= 0.337(3)$ and $c_{\nicefrac{1}{2}} = 0.27(4)$.

\begin{figure}
    \centering
    \includegraphics[width = 0.75\columnwidth]{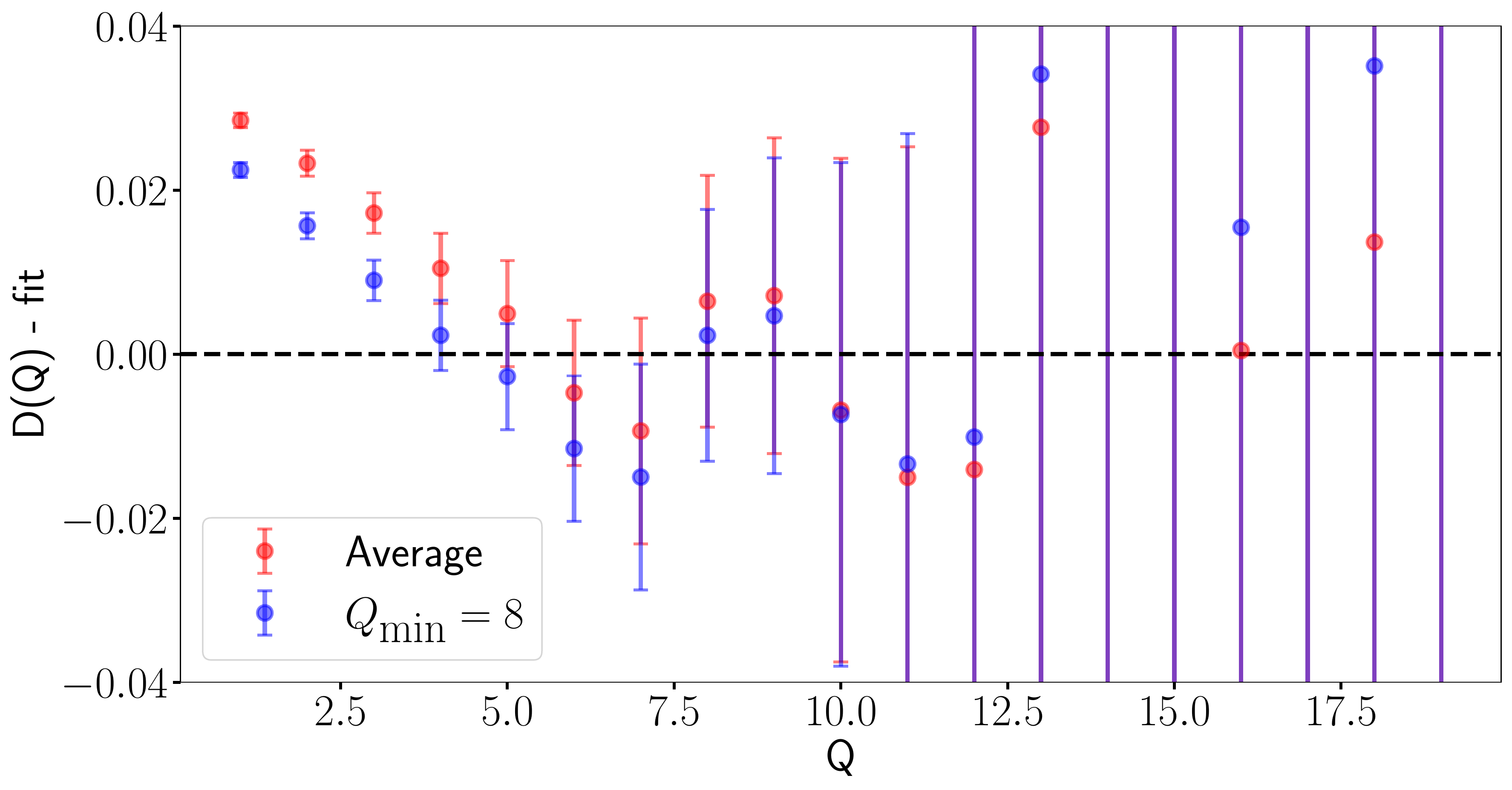}
    \caption{
    Difference between the data and the best-fit curves with the averaged parameters in eq.~\eqref{eq:results conformal dimension} (red) and with $Q_\text{min}=8$ (blue). For large values of $Q$ the uncertainties exceed the range displayed in the plot.
  }
    \label{fig: difference}
\end{figure}
In fig.~\ref{fig: difference} we plot the difference between the numerical data and the best-fit curve obtained using the averaged parameters. In the same plot, we also show the deviation for the best-fit curve obtained setting $Q_\text{min}=8$. For small charges, there are systematic deviations, but they become smaller than the numerical uncertainties for $Q\gtrsim 4 $. This justifies a posteriori the choice of fitting in the range $Q_{\rm{min}}\in[4,8]$. It is remarkable that also for small charges the relative deviations are rather small. For instance, the best-fit curve obtained for $Q_\text{min} = 8$ extrapolated to $Q=1$ agrees with the measurement within a 4\% relative error.

Notice that we did not try to fit the value of $c_0$, which we held fixed at its theoretical value $c_0\simeq -0.0937$. Indeed, as we argue below, the data are compatible with this value, but the increasing numerical uncertainties with the charge make it impossible to obtain a reliable estimate for $c_0$ or other subleading coefficients.

To justify the compatibility of $c_0=-0.0937$ with the Monte Carlo data, it is convenient to define the following quantity
\begin{equation}\label{eq:A(Q)}
\mathcal{A}(Q)=\mathcal{N}^{-1}(Q)\left[2 \frac{D(1+Q)}{\sqrt{1+Q}}-\frac{D(2+Q)}{\sqrt{2+Q}}
-\frac{D(Q)}{\sqrt{Q}}\right]\,,
\end{equation}
where
\begin{equation}
\mathcal{N}(Q)=\frac{2}{\sqrt{1+Q}}-\frac{1}{\sqrt{Q+2}}-\frac{1}{\sqrt{Q}}\,.
\end{equation}
The quantity $\mathcal{A}(Q)$ is so defined to be independent of $c_{{\nicefrac{3}{2}}}$ and $c_{\frac{1}{2}}$ when evaluated using eq.~\eqref{eq:conformal dimension of charged scalars}. Its $1/Q$ expansion reads\footnote{Here we restored two-subleading orders in the expansion of the scaling dimension $D(Q)$:
\begin{equation}\label{Eq_conformal_dimension_subleading}
D\left(Q\right) =c_{{\nicefrac{3}{2}}}Q^{3/2}+c_{\frac{1}{2}}Q^{1/2}+c_{0}+c_{-\frac12} Q^{-1/2}+c_{-1} Q^{-1}+O\left(Q^{-3/2}\right)\,.
\end{equation}
The $Q^{-1}$ term in this expression represents the contribution to the Casimir energy from higher derivative corrections to the Goldstone dispersion relation, see \cite{Cuomo:2020rgt} for details.} 
\begin{equation}\label{eq:A_expansion}
\mathcal{A}(Q)=c_0+\frac83\frac{c_{-\frac12}}{\sqrt{Q}}+5\frac{ c_{-1}}{Q}+O\left(Q^{-3/2}\right)\,.
\end{equation}
A similar sum rule was introduced in \cite{Hellerman:2015nra}.
\begin{figure}[t]
    \centering
    \includegraphics[width = 0.75\columnwidth]{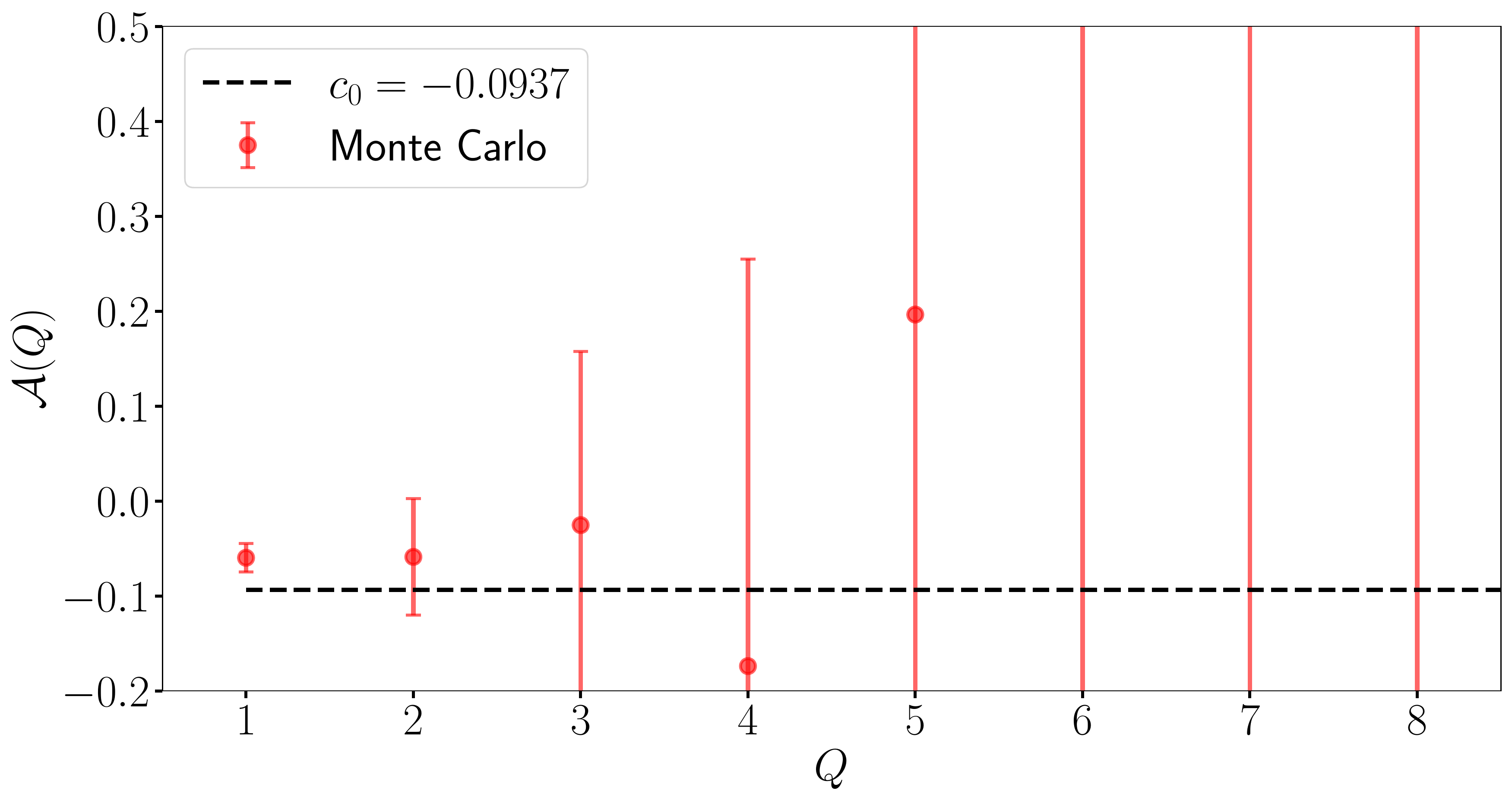}
    \caption{Results for $\mathcal{A}(Q)$, defined in eq.~\ref{eq:A(Q)}. Notice the large uncertainties despite the precision of the measurements of $\Delta(Q)$.}
    \label{fig:A(Q)}
\end{figure}

The results for $\mathcal{A}(Q)$ are presented in fig.~\ref{fig:A(Q)}. On the one hand, measurements for large charges do not achieve a sufficient level of precision to extract $c_0$. On the other hand, the results for small charges, for which the precision is high, are subject to unknown theoretical errors.\footnote{Perhaps relatedly, the extrapolation of the large $N$ analysis of \cite{Dondi:2021buw} suggests that non-perturbative terms in the series \eqref{eq:conformal dimension of charged scalars} are of the same order of $c_0$ for $Q\sim O(1)$.} 
Our results are nonetheless compatible with the theoretical value. %

\section{OPE coefficients}\label{sec:OPE coefficients}

Let us now discuss how to measure the OPE coefficients \eqref{eq_OPE_def}. We consider in particular the OPE coefficient in regime I (cfr. eq.~\eqref{eq:regime I asymptotic behavior}) for $Q_1=Q_2 = Q$, for which the EFT prediction reads
\begin{equation}\label{eq_OPE_I}
 \lambda_{Q,Q,-2Q}=\exp\left\{\frac{f\left(1\right)}{2\sqrt{\pi}}\left[c_{{\nicefrac{3}{2}}}Q^{3/2}+\alpha_{\nicefrac12}Q^{1/2}+\alpha_{0}+O\left(Q^{-1/2}\right)\right]\right\}\,.   
\end{equation}
where $f(1)\simeq 1$ and we included the first two subleading terms, which are multiplied by two unknown coefficients $\alpha_{\nicefrac12}$ and $\alpha_0$, for future reference.\footnote{The $Q^{1/2}$ term depends upon a subleading Wilson coefficient of the EFT which does not contribute to the scaling dimension $D(Q)$; therefore we cannot compute its value from the estimates obtained in sec.~\ref{sec:conformal dimension of scalar operators}. The $Q^0$ term, $\alpha_{0}$, is instead independent of Wilson coefficients, analogously to the $c_0$ term in eq.~\eqref{eq:conformal dimension of charged scalars}. In principle, its value could be computed  from the one-loop fluctuation determinant around the saddle-point of \cite{Cuomo:2021ygt}. In practice, this calculation is technically challenging and we treat $\alpha_0$ as an unknown parameter.} For the OPE in regime II, we take $Q_1=Q$ and $Q_2=1$. The theoretical prediction \eqref{eq:regime II asymptotic behavior} takes the form
\begin{equation}\label{eq_OPE_II}
\lambda_{1,Q,-Q-1}=C\left(1\right)Q^{D(1)/2}\left[1+\frac{0.46c_{{\nicefrac{3}{2}}}}{\sqrt{Q}}+\dfrac{\beta_{-1}}{Q}+O\left(Q^{-3/2}\right)\right]\,,
\end{equation}
where we also included an extra-subleading term, which depends upon a new Wilson coefficient $\beta_{-1}$.\footnote{This coefficient represents a subleading contribution in the operator matching \cite{Cuomo:2020rgt}.} For the sake of concreteness, in the following we discuss how to measure the OPE coefficient~\eqref{eq_OPE_I}. A similar discussion applies to the OPE in regime II.

In Monte Carlo simulations operators are normalized differently than in the CFT literature.
In Monte Carlo, 2pt functions at coincident points are normalized to 1, while in the CFT literature 2pt functions are normalized
to $1/|x|^{2 D(Q)}$ asymptotically. Therefore, to extract the OPE coefficient as defined in eq.~\eqref{eq_OPE_def}, 
we measure a suitable ratio between the 3pt function and 2pt functions. The ratio of interest is
\begin{align}
\dfrac{\left\langle \mathcal{O}_{-Q}\left(x\right)\mathcal{O}_{2Q}\left(0\right)\mathcal{O}_{-Q}\left(-x\right)\right\rangle }{\sqrt{\left\langle \mathcal{O}_{2Q}\left(0\right)\mathcal{O}_{-2Q}\left(x\right)\right\rangle }\left\langle \mathcal{O}_{Q}\left(0\right)\mathcal{O}_{-Q}\left(x\right)\right\rangle }  = 2^{D(2Q)-2D(Q)}\lambda_{Q,Q,-2Q}\,,\label{eq:OPE coefficients large charge function}
\end{align}
where on the right-hand side we expressed it in terms of the OPE coefficient~\eqref{eq_OPE_def}, using the continuum infinite size CFT prediction.

\begin{figure}[t]
    \centering
    \includegraphics[width=0.75\columnwidth]{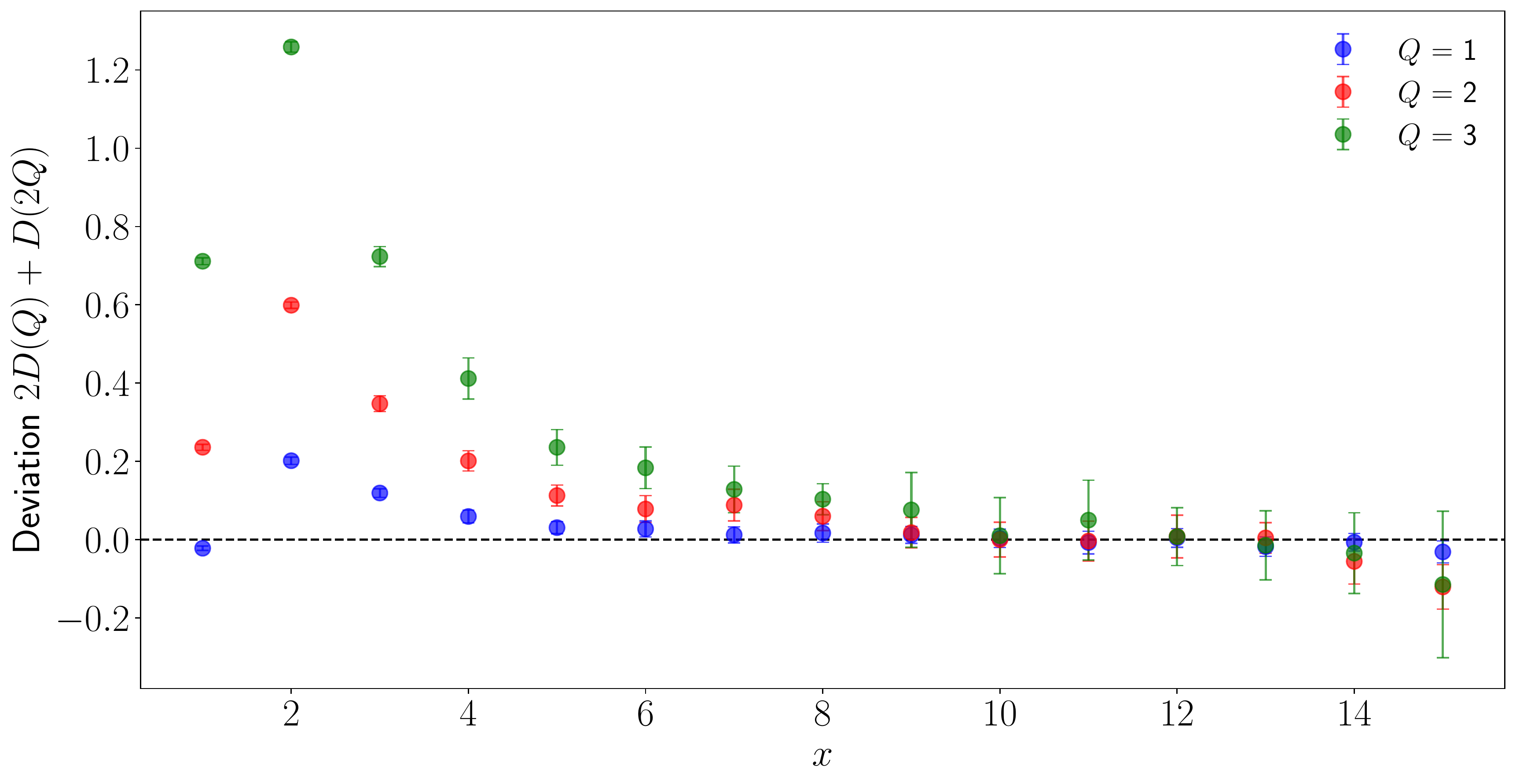}
    \caption{Difference between the value of $\gamma_Q$ measured from Monte-Carlo and the theoretical prediction $2D(Q) + D(2Q)$ (see eq.~\eqref{eq_ratio_OPE}). The result is obtained from the ratio of correlation functions sampled at $L=32$ and $L=64$, at the same relative position. 
    We used the values $D(Q)$ and $D(2Q)$ measured in the previous section for the theoretical prediction.}
    \label{fig:Ratio 3pt functions in the torus}
\end{figure}

In order to measure 3pt functions with the Worm algorithm, we need to rewrite them as some combination of 2pt functions in the presence of background charges, as in eq.~\eqref{eq_Rq_MC}.\footnote{This is because the Worm algorithm can only generate configurations with 2 charge insertions, corresponding to its tail and head.} We, therefore, write the OPE coefficient as:
\begin{equation}\label{eq:continuum infinite size predictions}
    \left\langle \mathcal{O}_{-Q}\left(x\right)\mathcal{O}_{2Q}\left(0\right)\mathcal{O}_{-Q}\left(-x\right)\right\rangle=\left\langle \mathcal{O}_{-Q}\left(x\right)\mathcal{O}_{Q}\left(0\right)\right\rangle_{Q_{0}-Q_{-x}}\left\langle \mathcal{O}_{Q}\left(0\right)\mathcal{O}_{-Q}\left(-x\right)\right\rangle.
\end{equation}
The task of measuring OPE coefficients reduces to the measurement of 2pt functions in the presence of background charges; these can be efficiently sampled in terms of ratios (as in eq.~\eqref{eq:decomposition 2pt function in ratios}) using the strategy outlined in the previous section. Further details are given in App.~\ref{sec:ratio between correlation functions.}.

We now discuss lattice and finite-size effects. First, it is useful to determine the region where lattice effects are negligible. To this aim, we consider the ratio of three-point functions at different lattice sizes but at the same relative position:
\begin{equation}\label{eq_ratio_OPE}
\frac{T_{Q,L}(x)}{T_{Q,\alpha L}(\alpha x)}=\alpha^{\gamma_Q}\,,\qquad
\gamma_Q=D(2 Q)+2 D(Q)\,,
\end{equation}
where we defined the 3pt function on the lattice as
\begin{equation}
T_{Q,L}(x)=\left\langle \mathcal{O}_{-Q}\left(x\right)\mathcal{O}_{2Q}\left(0\right)\mathcal{O}_{-Q}\left(-x\right)\right\rangle_{T^3_L}\,;
\end{equation}
the value for the exponent $\gamma_Q$ in eq.~\eqref{eq_ratio_OPE} follows from scale invariance in the CFT. Analogously to eq.~\eqref{eq:ratios}, the ratio \eqref{eq_ratio_OPE} is independent of $x$ when lattice effects are negligible.

In fig.~\ref{fig:Ratio 3pt functions in the torus} we plot the difference between the exponent $\gamma_Q$ extracted from numerical measurements for different values of $x$ and the CFT prediction in eq.~\eqref{eq_ratio_OPE}. We show explicitly the results for $Q=1,2,3$ (the plot for $Q=4$ is analogous) and $L=32$. The plot clearly shows that the region where lattice effects are negligible decreases with the charge. This makes it challenging to perform measurements for large values of the charge.

\begin{figure}[t]
    \centering
    \includegraphics[width=0.75\columnwidth]{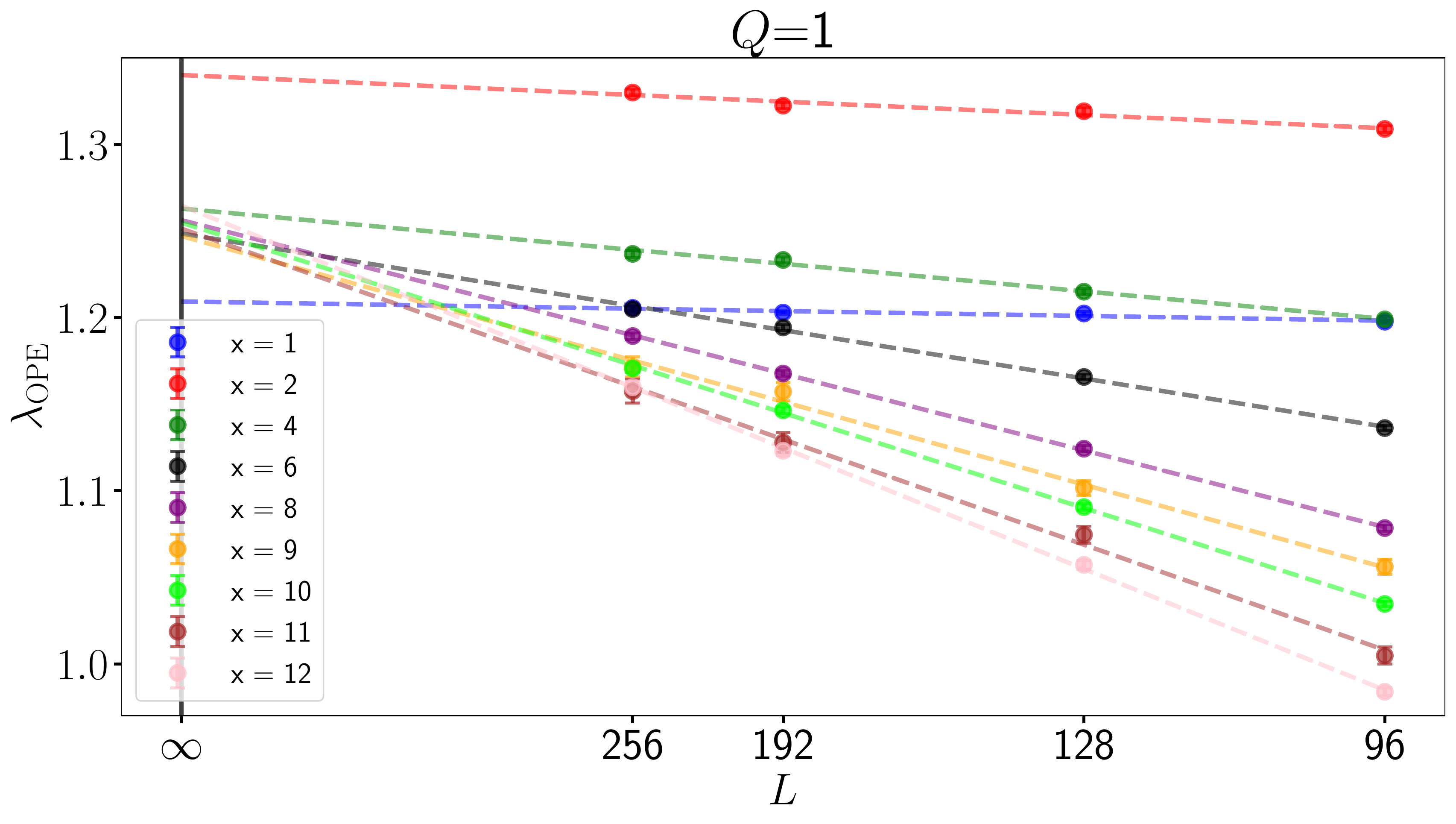}
    \caption{Linear extrapolation of the OPE coefficient $\lambda_{Q,Q,-2Q}$ to $L\to\infty$ at fixed $x$ and $Q=1$.}
    \label{fig:extrapolation}
\end{figure}

Differently than with 2pt functions, it is not possible to eliminate completely systematic errors from finite size effects.\footnote{Notice that ratios of three-point functions at different lattice size are independent of the OPE coefficient, as eq.~\eqref{eq_ratio_OPE} shows.}
Therefore, to extract the OPE coefficient we fix $x$ and study the $L$ dependence of the ratio of 3- and 2-pt functions in eq.~\eqref{eq:OPE coefficients large charge function}. The idea is that, as long as $x$ is outside the region where lattice effects are relevant, the extrapolation to $L\to\infty$ is unaffected by lattice corrections. Moreover, in the limit $x/L\to0$, at fixed $x$, the lattice 3pt function should be well described by the continuum infinite size prediction. Thus, this provides a direct measurement of the OPE coefficients. We show the results for $Q=1$ in fig.~\ref{fig:extrapolation}. While for $x\in[1,2,4,6]$ there are deviations, all results for $x>6$ converge to the same value, within uncertainties. The error bars for the final result are estimated from the dispersion of the intercept.

\begin{figure}[t]
\centering
\begin{subfigure}
    \centering
    \includegraphics[width=0.75\columnwidth]{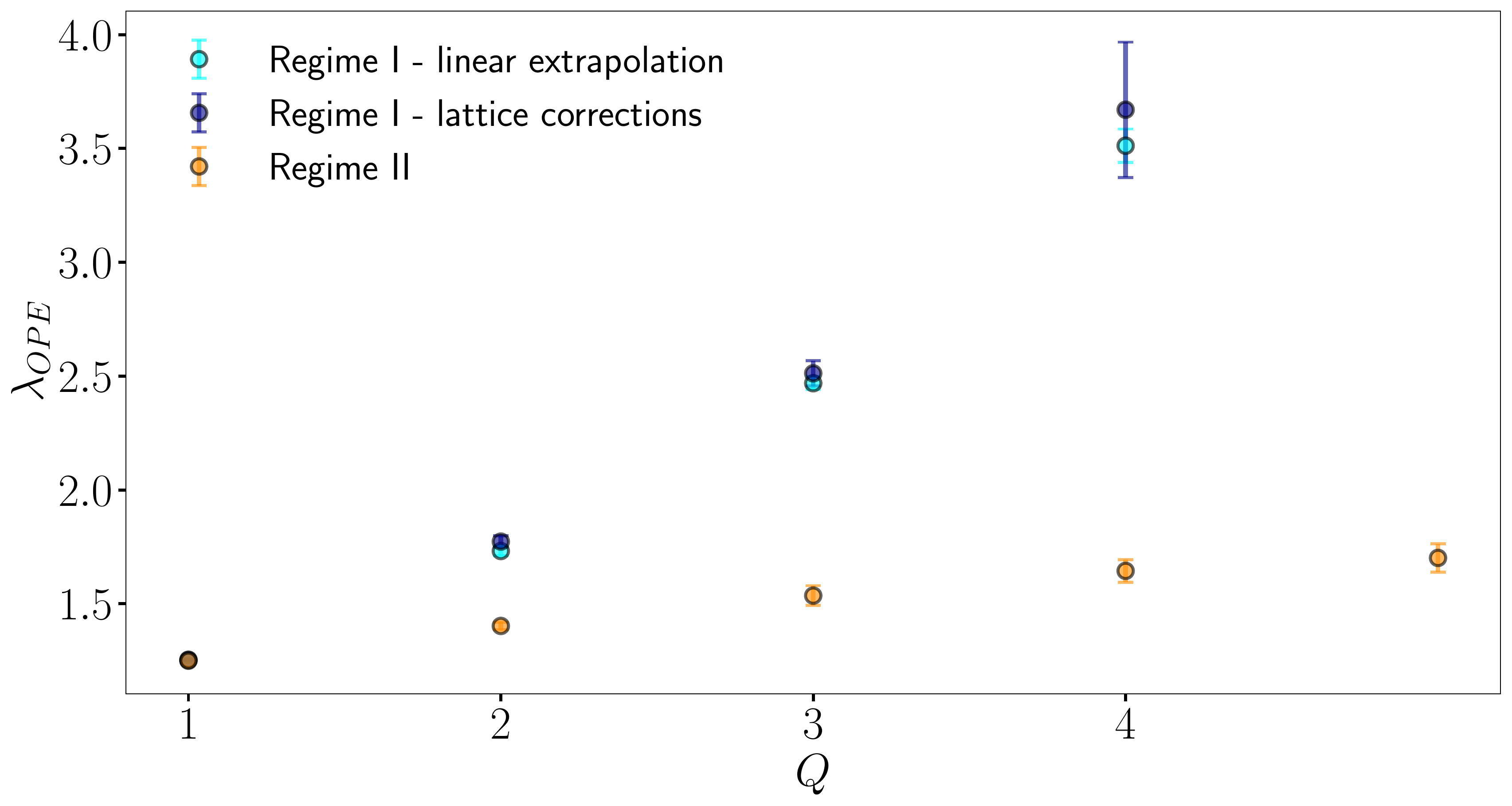}
    \caption{Numerical results. In regime I, the OPE coefficients extracted from the linear extrapolation are in cyan, and OPE coefficients extracted using lattice corrections are in dark blue. In regime II, both methods yield the same results. The OPE coefficient for $Q=1$ is the same in both regimes. Notice that the OPE coefficient in regime I grows much faster with $Q$ than the one in regime II; this is in qualitative agreement with the EFT predictions. The data in this plot is in the table below. }
    \label{fig: OPE raw}
\end{subfigure}
\begin{subtable}
    \centering
    \begin{tabular}{|c|c|c|c|c|c|}

\hline 
$Q$ & 1 & 2 & 3 & 4 & 5\tabularnewline
\hline 
\hline 
Regime I - linear extrapolation & 1.252(6) & 1.73(2) & 2.46(3) & 3.51(7) & -----\tabularnewline
\hline 
Regime I - lattice corrections & 1.254(2) & 1.77(3) & 2.51(6) & 3.7(3) & -----\tabularnewline
\hline 
Regime II & 1.250(9) & 1.40(2) & 1.54(4) & 1.64(5) & 1.70(6)\tabularnewline
\hline 
    \end{tabular}
\end{subtable}
\end{figure}

For larger values of $Q$ one obtains similar plots, but the data has larger uncertainties and the extrapolation for $L\to\infty$ does not converge as nicely. This is due to the increased significance of lattice effects, as shown in fig.~\ref{fig:Ratio 3pt functions in the torus}. To improve the precision and accuracy, we used conformal perturbation theory to parameterize both lattice corrections and finite size effects in eq.~\eqref{eq:OPE coefficients large charge function}. We obtain an expression as a series in powers of $a/x$, $a/L$ and $L/x$, see App.~\ref{sec:lattice corrections} for details. We improve on the naive linear extrapolation by fitting the coefficients of these powers. Notice that in this approach we perform a unique multidimensional-fit with all the data (i.e. for all values of $x$ and $L$), rather than performing separate linear extrapolations for each value of $x$.

The numerical values for the OPE coefficients in regimes I and II are shown in fig.~\ref{fig: OPE raw}. We show both the results of the linear extrapolation, as well as those obtained from accounting lattice and finite size corrections, as explained in the previous paragraph. Both methods yield the same results for the OPE coefficients Regime II. Also for the OPE coefficients in regime I the two measurements are compatible, but the results accounting for lattice corrections, in dark blue, are consistently larger than those obtained from the linear extrapolation, in cyan. This suggests that our results for $\lambda_{Q,Q,-2Q}$ may be underestimated. Unfortunately, to further investigate this issue we would need to perform more precise simulations at larger distances, which are currently beyond our reach.\footnote{Our results for $Q=4$ are obtained with $L_{max}=256$ and $x_{max}=12$.} Our results are tabulated in the table in fig.~\ref{fig: OPE raw}.

The range of charges that can be sampled in regime I is limited, as the statistical errors grow quickly with the charge. In regime II, it would be possible to go further, but this would not improve the precision with which we can measure the coefficient $c_{{\nicefrac{3}{2}}}$, as we will explain in the following paragraphs.

Let us now analyze our data. To this aim, for the OPE coefficient in \textbf{regime I}, we perform two separate fits: one for the first three coefficients in eq.~\eqref{eq_OPE_def} and all data points, and one for the first two Wilson coefficients only and the results for $Q\geq 2$. The results are shown in Tab.~\ref{tab:OPE_regime_I}.

\begin{table}[h]
    \centering
        \begin{tabular}{|c|c|c|c|c|}
        \hline 
         data & charges   &  $c_{{\nicefrac{3}{2}}}$ & $\alpha_{\nicefrac12}$ & $\alpha_{0}$ 
         \tabularnewline
        \hline
         lattice corrections
         & $Q\geq 1$   &  0.316(15) & 1.57(8) &  -1.09(7) \tabularnewline
         \hline
         linear extrapolation
         & $Q\geq 1$   &  0.33(4) & 1.4(2) &  -0.9(2) \tabularnewline
        \hline      lattice corrections     &
       $Q\geq 2$   &  0.446(8) & 0.176(7) &  ----\tabularnewline
        \hline      linear extrapolation     &
       $Q\geq 2$   &  0.43(4) & 0.17(3) &  ----\tabularnewline
        \hline
        \end{tabular}
    \caption{Result for the fits of the OPE coefficient in regime I with different free parameters and for different values of the charges in eq.~\ref{eq_OPE_I}.
    We show the fits with both the data obtained from the linear extrapolation and with the inclusion of lattice corrections; notice the latter are more precise.}
    \label{tab:OPE_regime_I}
\end{table}

\begin{figure}[t]
 \centering
    \includegraphics[width = 0.75\columnwidth]{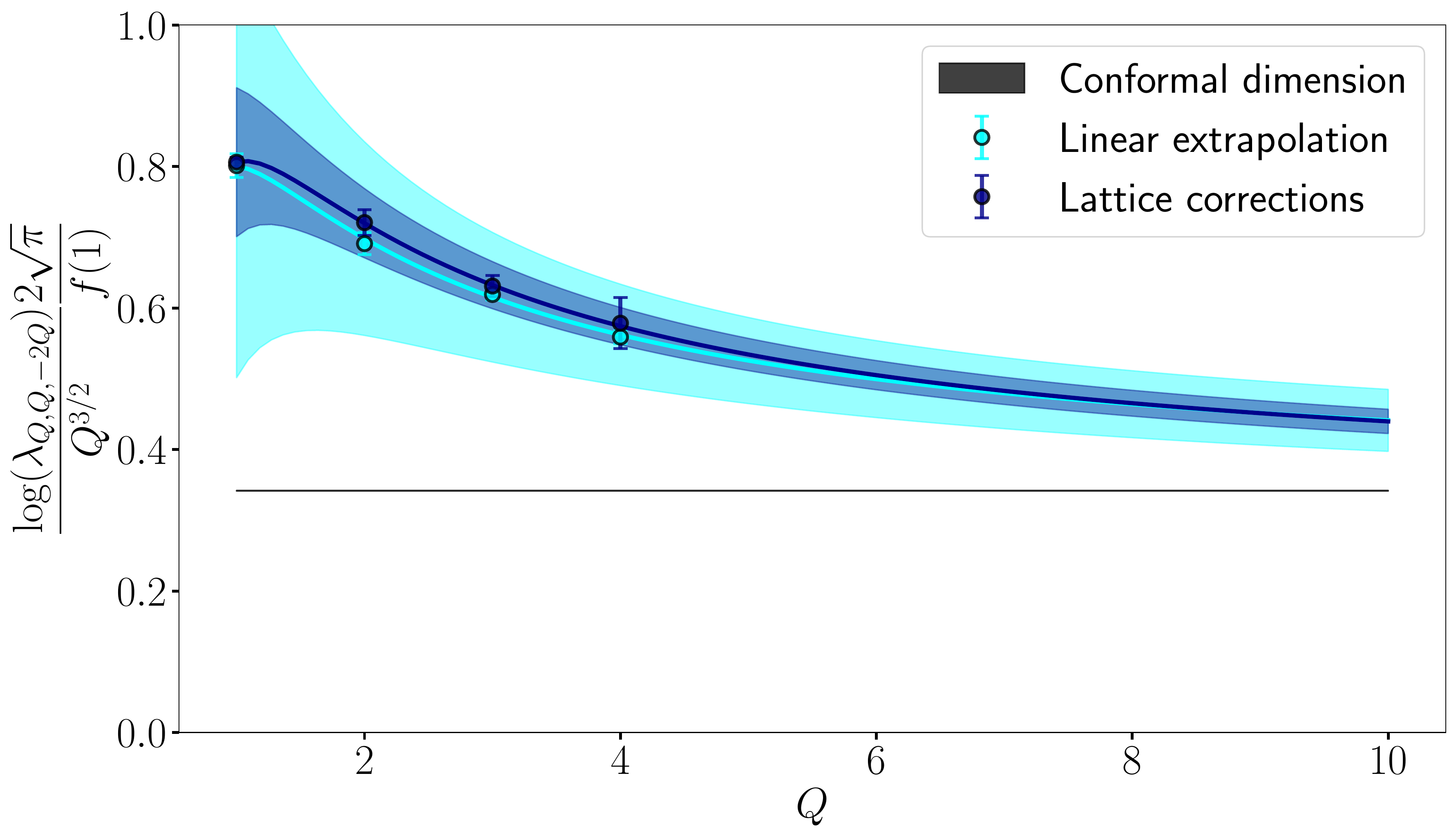}
  \centering
\captionof{figure}{Comparison between the numerical results for eq.~\eqref{eq_log_OPE_I} and the predicted value $c_{{\nicefrac{3}{2}}}=0.34(1)$. The black line is the value of $c_{{\nicefrac{3}{2}}}$ obtained in sec.\ref{sec:conformal dimension of scalar operators}, and its width is the uncertainty. The coloured full lines are the fits and the coloured regions around their uncertainty. The fits converge to the black line for larger values of $Q$.}
    \label{fig:log OPE}
\end{figure}

Clearly, our analysis is limited by the small number of data points. Nonetheless, the estimates in table~\ref{tab:OPE_regime_I} suggest that the leading Wilson coefficient should lie in the range $c_{{\nicefrac{3}{2}}}\approx 0.4\pm0.1$, which is compatible with the value $c_{{\nicefrac{3}{2}}}\approx 0.34$ obtained from the measurements of scaling dimensions in the previous section. To appreciate this point better, in
fig.~\ref{fig:log OPE} we show our results for
\begin{equation}\label{eq_log_OPE_I}
\dfrac{\log(\lambda_{Q,Q,-2Q})}{Q^{3/2}} \dfrac{2 \sqrt{\pi}}{f(1)} \sim c_{{\nicefrac{3}{2}}} + O(Q^{-1})\,.
\end{equation}
The value of $c_{{\nicefrac{3}{2}}}$ can be extracted from the asymptotic behaviour of this quantity. The value obtained with the linear extrapolation is smaller than the one with lattice corrections. The range of charges is small hence it is difficult to extrapolate to infinite charge. Nevertheless, the results are compatible with asymptotically approaching $c_{3/2}\approx 0.34$.

\begin{figure}[t]
\centering
\begin{minipage}{.5\textwidth}
    \centering
    \includegraphics[width=\columnwidth]{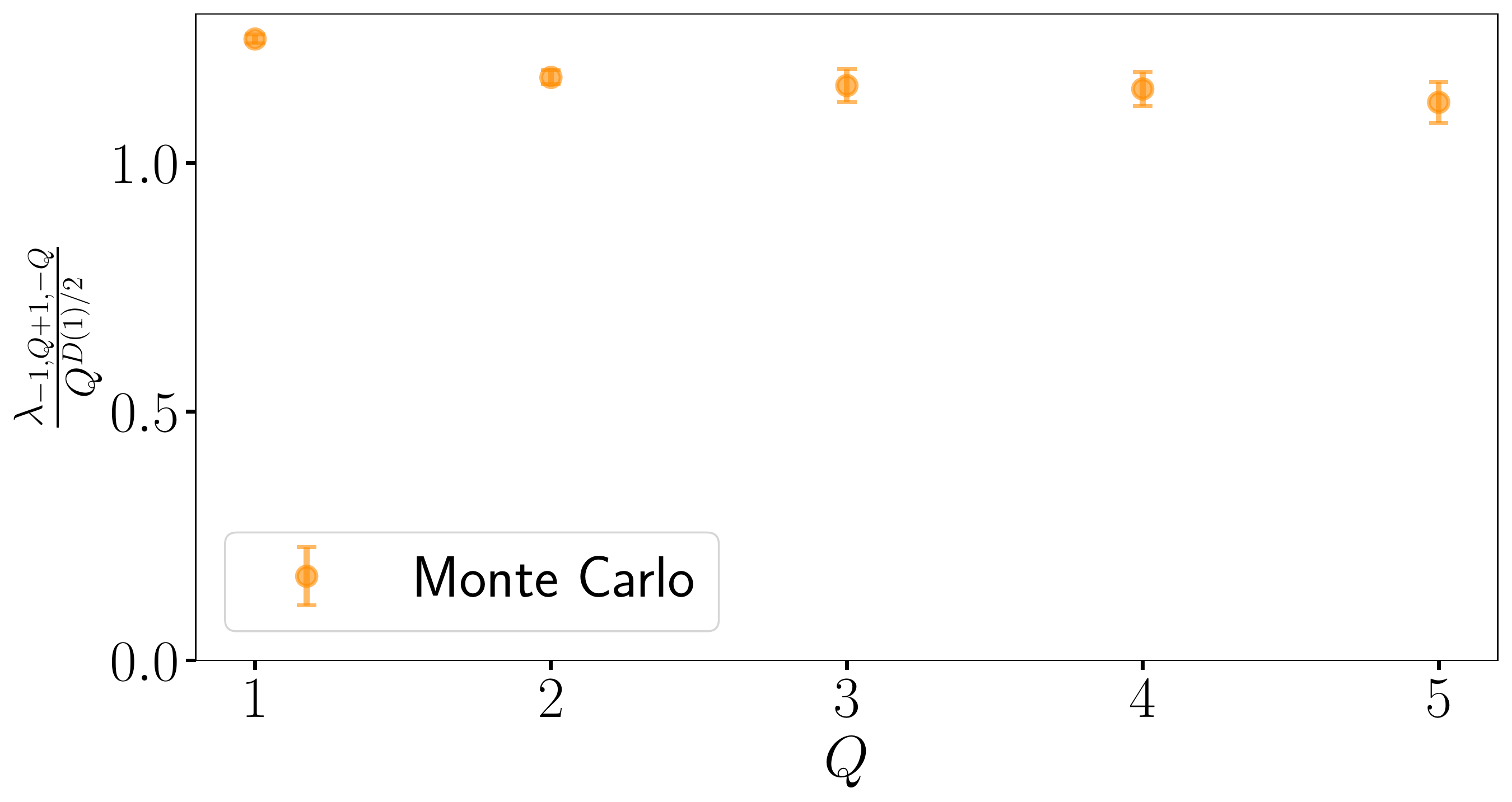}

\end{minipage}%
\begin{minipage}{.5\textwidth}
    \centering
    \includegraphics[width = \columnwidth]{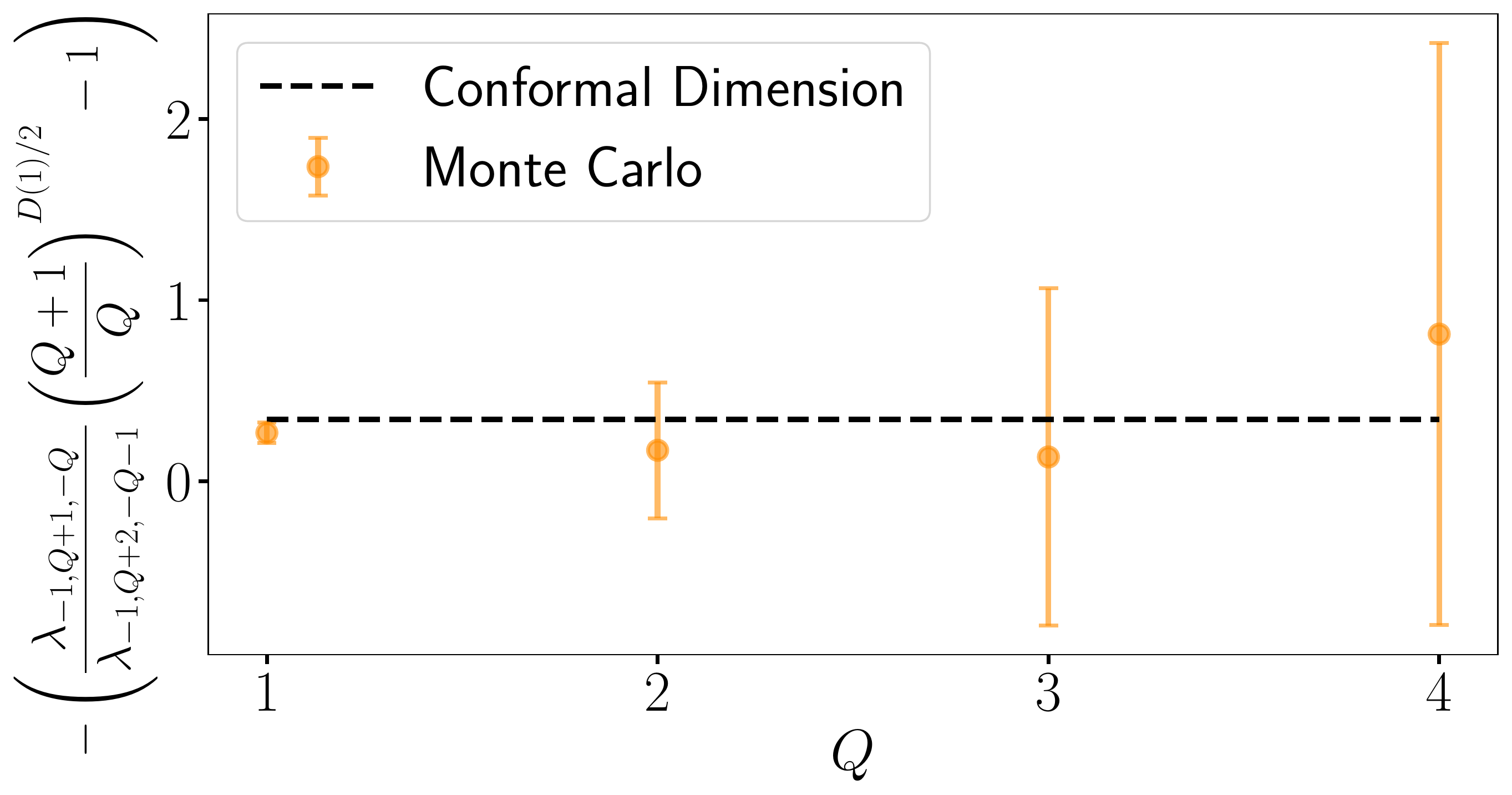}

\end{minipage}
 \par
\medskip
\noindent
\begin{minipage}[t]{.49\textwidth}
  \centering
\captionof{figure}{Coefficient of leading behaviour OPE coefficient in regime II.}
\label{fig:leading behavior regime II}
\end{minipage}%
\hfill
\begin{minipage}[t]{.49\textwidth}
    \centering
    \captionof{figure}{Ratio in eq.\eqref{eq:remove normalization}. The black line is evaluation with $c_{{\nicefrac{3}{2}}}=0.339(1)$, obtained in sec.\ref{sec:conformal dimension of scalar operators}.}
    \label{fig:lambda 1 and lambda2}
\end{minipage}
\end{figure}

We now discuss the results for the OPE coefficient in \textbf{regime II}. 
We begin by testing the leading behaviour of the OPE coefficient. In fig.~\eqref{fig:leading behavior regime II}, we show $\lambda_{1,Q,-Q-1}/Q^{D(1)/2}$. If the EFT prediction eq.~\eqref{eq_OPE_II} holds, this ratio should approach a constant for large $Q$. The plot shows that this is indeed the case.

Having checked the leading behaviour, we focus on studying the sub-leading correction and measuring $c_{{\nicefrac{3}{2}}}$. Unfortunately, fits are not very precise due to large correlations between $C(1)$ and $c_{{\nicefrac{3}{2}}}$ in the range of charges available. For reference, the fits including and excluding the subleading coefficient $\beta_{-1}$ (cfr. eq.~\eqref{eq_OPE_II}) are given in table \ref{tab:C1}.
\begin{table}
\centering
\begin{tabular}{|c|c|c|}
        \hline 
           $C(1)$ & $c_{{\nicefrac{3}{2}}}$ & $\beta_{-1}$ 
         \tabularnewline
        \hline
           1.14(6) & -0.2(3) &  0.18(7) \tabularnewline
         \hline
          1.02(2) & 0.47(5) &  ---- \tabularnewline
       \hline
        \end{tabular}
        \caption{Result for the fit of the OPE coefficient in Regime II, with different parameters, in eq.~\eqref{eq_OPE_II}. }
        \label{tab:C1}
\end{table}
Clearly the fit including the coefficient $\beta_{-1}$ makes the uncertainties too large for the results to be meaningful. The second fit is compatible within $2\sigma$ with the estimate $c_{{\nicefrac{3}{2}}}\approx 0.34$.

To study directly the first sub-leading contribution in eq.~\eqref{eq_OPE_II},
we consider the following ratio
\begin{equation}\label{eq:remove normalization}
   -\dfrac{Q^{3/2}}{0.23}\left(\dfrac{\lambda_{1,Q+1,-Q-2}}{\lambda_{1,Q,-Q-1}}\left(\dfrac{Q}{Q+1}\right)^{D(1)/2} - 1\right)\sim c_{{\nicefrac{3}{2}}}  +O(Q^{-1/2})\,.
\end{equation}
The EFT predicts that eq.~\eqref{eq:remove normalization} asymptotes to $c_{{\nicefrac{3}{2}}}$ for large $Q$. Our numerical results are shown in fig.~\ref{fig:lambda 1 and lambda2}. They are compatible with the value of $c_{{\nicefrac{3}{2}}}$ obtained in sec.~\ref{sec:conformal dimension of scalar operators}, represented by the black dashed line. However, similarly to the analysis $c_0$ in the previous section, the uncertainties increase rapidly with $Q$, making it impossible to obtain a reliable estimate. These uncertainties also represent an obstacle towards improving our results with measurements of the OPE coefficients at larger values of $Q$.

\section{Conclusions and outlook}\label{sec:conclusions}

In this work, we used Monte-Carlo calculations to test the validity of the superfluid EFT for describing the large charge sector of the $O(2)$ model. Our results were already summarized in the introduction. Here we instead discuss the implications of our findings and potential future directions.

The most surprising aspect of the result for $D(Q)$ is the effectiveness of the large charge predictions also for $Q=O(1)$. For instance, the extrapolation of the best-fit curve obtained from the data with $Q\geq 8$ reproduces the measured value of $D(1)$ with 2\% accuracy. Overall, our results confirm that this phenomenon persists also for OPE coefficients.  A partial justification for the accuracy of the large charge expansion for the scaling dimension was given in \cite{Dondi:2021buw} via resurgence analysis of the large $N$ result of \cite{Alvarez-Gaume:2019biu}. It might be interesting to perform similar analyses for OPE coefficients.

While our results are encouraging, the scarcity of data points does not allow us to draw unambiguous conclusions about the validity of the superfluid EFT. It is therefore important to obtain more data. Unfortunately, obtaining OPE coefficients for higher values of the charges is beyond the reach of our current Monte-Carlo algorithm. For instance, we estimate that obtaining the OPE coefficient in Regime I for $Q_1=Q_2=5$, with an uncertainty of 10\%, would require 10 CPU years. 

A simpler target might be the Monte-Carlo calculation of the scaling dimension of the lightest charged operator with spin $J=2$.\footnote{We focus on spin 2 since the lightest charged operator with spin 1 is expected to be a descendant of the scalar operator $\mO_Q$.} Notice that the cubic symmetry group of the lattice naturally allows to represent operators up to spin $J<4$.\footnote{We thank Luca Delacr\'etaz for useful discussions on this.}
As commented in the introduction the superfluid EFT predicts the scaling dimension of the spin $2$ operator to be $D(Q)+\sqrt{3}$. However, it is unclear whether one should expect this result to converge for small values of $Q$, as it happens for scalar operators. Indeed it is expected that the large charge sector of the $O(2)$ model admits a rich phase diagram as a function of the ratio $J/Q$ \cite{Cuomo:2017vzg,Cuomo:2022kio}. We hope to report about progress in this direction in the future.

It would also be interesting to explore alternative methods to compute the spectrum of charged operators in the $O(2)$ model. An intriguing possibility is provided by the fuzzy-sphere regularization of \cite{Zhu:2022gjc}, which allows directly computing the spectrum of the theory on the cylinder.\footnote{We thank Andreas Läuchli for discussions regarding his ongoing work in this direction.}  We were also informed of interesting numerical bootstrap results for charged operators in the $O(3)$ model \cite{Ning_informal}.

The accuracy of the large charge expansion is reminiscent of the success of the Regge relation for the QCD and Yang-Mills spectrum \cite{Gribov:2003nw}, and more recently of the remarkable results of the large spin bootstrap in the $O(N)$ models \cite{Simmons-Duffin:2016wlq,Liu:2020tpf}.
In both of these examples, the results are (partially) explained by the analyticity properties of the spectrum \cite{Caron-Huot:2017vep,Correia:2020xtr,Caron-Huot:2020ouj}. It remains an important open question whether the CFT data of the $O(2)$ model enjoy similar analyticity properties as a function of the charge. %

\acknowledgments

GC is supported by the Simons Foundation (Simons Collaboration on the Non-perturbative Bootstrap) grants 488647 and 397411. 
JM is supported by FCT with the fellowship 2021.04743.BD, co-funded by the Programme Por$\_$Norte, the European Social Fund (ESF), and the Portuguese state budget (MCTES).
JM, JO and JV thank the cluster time provided by INCD funded by FCT and FEDER under project 01/SAICT/2016 nº 022153 and the grant 2021.09830.CPCA of the Advanced Computing Projects (2nd edition) as well as GRID FEUP. They also thank Centro de Física do Porto funded by Portuguese Foundation for Science and Technology (FCT) within the Strategic Funding UIDB/04650/2020.
JP is supported by the Simons
Foundation grant 488649 (Simons Collaboration on the Nonperturbative Bootstrap) and
the Swiss National Science Foundation through the project 200020\_197160 and through
the National Centre of Competence in Research SwissMAP.

\appendix
\section{Monte Carlo}\label{sec:Monte Carlo}
This section describes the Monte Carlo method and measurement strategies employed. It describes the world line formulation, the Worm algorithm and the procedures required to express the correlation functions as averages that can be efficiently estimated with Monte Carlo. We describe an improved Worm update, the continuous time update, that guarantees the Worm tail always moves. 

The lattice Hamiltonian of the O(2) model is 
\begin{equation}
H=-\beta \sum_{n, \rho} \cos \left(\theta_n-\theta_{n+a \hat{\rho}}\right), 
\end{equation}
where the field $\theta_x$ is defined at the cubic lattice nodes. Simulations can be performed in this representation. However, it is more efficient to use a world-line representation \cite{Banerjee:2010kc}, where the node variables are mapped into edge variables using
\begin{equation}
\exp \{\beta \cos \left(\theta_n-\theta_{n+a \hat{\rho}}\right)\}=\sum_{k=-\infty}^{\infty} I_k(\beta) e^{i k \left(\theta_n-\theta_{n+a \hat{\rho}} \right)},
\end{equation}
where $I_k(\beta)$ is the modified Bessel function of the first kind and $\beta$ is the  inverse temperature. We work at the critical temperature of the three dimensional $O(2)$ model, $\beta=0.4541652$ \cite{ballesterosFiniteSizeEffects1996}. Since each $k$ is associated with a pair $\left(\theta_x,\theta_{x+a \hat{\rho}}\right)$, these live on the edge of the lattice connecting $x$ to $x+a \hat{\rho}$. After this rewriting, the path integral over $\theta$ can be performed explicitly and the partition function becomes a sum over all possible values $k$ for all the edges 
\begin{equation}
Z=\sum_{\{k\}} \prod_{n, \hat{\rho}\in\{\hat{1},\hat{2},\hat{3}\}}\left\{I_{k_{n, n + a \hat{\rho}}}(\beta)\right\} \delta\left(\sum_{\hat{\rho}}\left(k_{n, n + a\hat{\rho}}-k_{n-a\hat{\rho}, n}\right)\right),
\end{equation}
where the sum is over all possible configurations of edge variables and the product is over nodes, $n$, and the edges connected to it.

The world-line formulation brings two significant improvements. The first is the possibility of using the Worm algorithm \cite{Prokofev:2001ddj}, which has one of the smallest dynamical critical exponents \cite{Prokof_ev_2001}. The second is that correlation  functions can be reinterpreted as the partition function in the presence of some background charge
\begin{equation}
\left\langle\prod_i \exp \left(Q_i \delta_{i x}\right)\right\rangle= \left\langle 1 \right\rangle_{\sum_i\left(Q_i\right)_{x_i}},
\end{equation}
where the subscript $\sum_i\left(Q_i\right)_{x_i}$ indicates that the correlation function is computed with the partition function $Z_{\sum_i\left(Q_i\right)_{x_i}}$
\begin{equation}
Z_{\sum_i\left(Q_i\right)_{x_i}}=\sum_{\{k\}} \prod_{n, \hat{\rho}}\left\{I_{k_{n, n+ a\hat{\rho}}}(\beta)\right\} \delta\left(\sum_{\hat{\rho}}\left(k_{n, n+a\hat{\rho}}-k_{n-a\hat{\rho}, n}\right)+\sum_i Q_i \delta_{i n}\right).
\end{equation}

Since $\sum_{\hat{\rho}}\left(k_{n, n+a\hat{\rho}}-k_{n-a\hat{\rho}, n}\right)$ is interpreted as charge conservation at each node, the extra term $\sum_i\left(Q_i\right)_{x_i}$ can be interpreted as a source/sink of charge, or in other words, a background charge distribution. 

\subsection{Worm algorithm}

The Worm algorithm consists of
two steps. An update step generates new configurations, Alg.~\ref{alg:Update-step.-This}. Then, a measurement step %
extracts the desired correlation function, %
see Alg.~\ref{alg:Measurement-step:-this}.
\begin{algorithm}
\SetAlgorithmName{Algorithm}

\text{1.~}Pick~a~random~site~$x_{h}$~(head~site).~Define~$x_{t}=x_{h}$~(tail~site).

\text{2.~}Randomly~pick~a~direction~$\hat{\rho} \in \{\hat{1}, \hat{2},\hat{3}\}$ and an orientation $\sigma =\pm 1$.

\text{3.~}Let~$k$~be~the~current~flowing~through~the~bond~connecting~$x_{t}$~to~$x_{t}+\sigma \hat{\rho}$.~
If:
\begin{itemize}
\item $\sigma=1$ update $k\to k+1$ with probability $I_{k+1}\left(\beta\right)/I_{k}\left(\beta\right)$;
\item $\sigma=-1$   update $k\to k-1$ with probability $I_{k-1}\left(\beta\right)/I_{k}\left(\beta\right).$
\end{itemize}
\text{4.}~If:
\begin{itemize}
\item update is accepted: $x_{t}=x_{t}+\sigma\hat{\rho};$
\item update is not accepted: $x_{t}=x_{t}.$
\end{itemize}
\text{5.~}If~$x_{t}=x_{h}$~the~update~ends.~Else~go~to~step~2.~
\caption{Update step.\label{alg:Update-step.-This}}
\end{algorithm}

\begin{algorithm}
\SetAlgorithmName{Algorithm}{}
\text{1.~}Start~two~counters~$c_x=0$ and $c_y=0$, that count the number of times the head and the tail coincide and the number of times the tail is at $y$, respectively.~

\text{2.~}Define~the~head~and~tail~sites~as~$x$:~$x_{t}=x_{h}=x.$~

\text{3.~}Perform~step~2,~3~and~4~of~the~update~step.

\text{4.~}Whenever~$x_{t}=y$,~increment~the~counter~$c_y=c_y+1.$~

\text{5.~}When~$x_{t}=x$,~increment~the~counter~$c_x=c_x+1$.~Otherwise~go~back~to~step~3.~

\text{6.~}Repeat the previous steps $N$ times. The expectation value is given by $\nicefrac{c_y}{c_x}.$ 
\caption{Measurement step: $\left\langle e^{i\theta(x)}e^{-i\theta(y)}\right\rangle$.
\label{alg:Measurement-step:-this}}
\end{algorithm}

In the measurement step, the head of the Worm transports charge 1, and it generates configurations with charge insertions at its head and tail. Then the ratio between the number of times the head and tail are at the positions at which the correlation function is being measured and the number of times the head and the tail are at the same position is an estimate of the correlation function. 

\begin{figure}[t]
    \centering
    \includegraphics[width = 0.75\columnwidth]{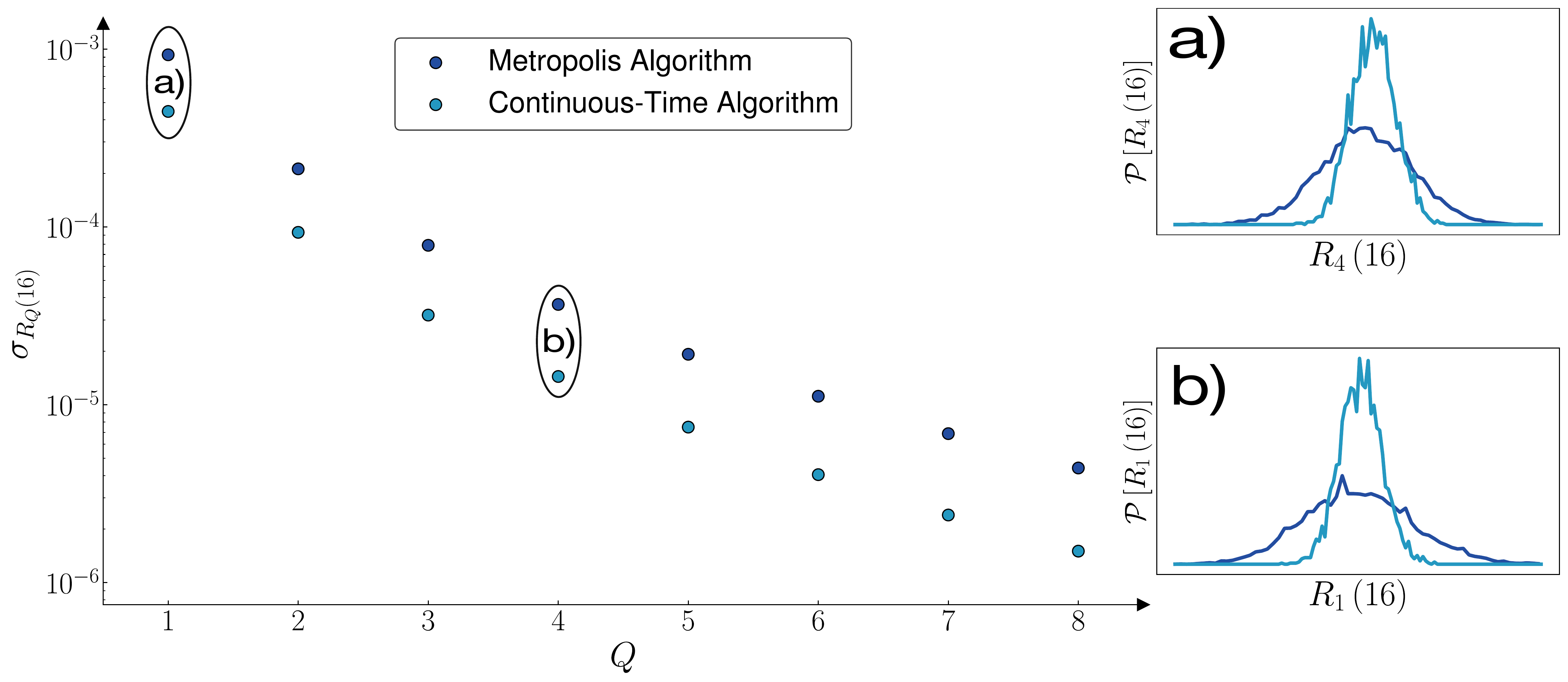}
    \caption{Variance of $R_Q(16)$ as a function of $Q$, defined in eq. \eqref{eq:decomposition 2pt function in ratios}. The comparison is made at a fixed CPU time. The histograms show the distribution of $R_Q(16)$, as obtained from the measurement step described in alg. \ref{alg:Measurement-step:-this}.}
    \label{fig:comparisonCT_vs_M}
\end{figure}

This algorithm can be improved by modifying step 3 and removing step 4. Instead of uniformly choosing a direction and then choosing to accept it, we can immediately move in a given direction with a probability that makes it equivalent to choosing a direction and then accepting or not that direction. We denote this algorithm by \textbf{continuous time update}, contrary to the Metropolis-like update of the original algorithm. This is achieved by choosing to move through a given edge with probability $P(\sigma\hat{\rho})$, given by  
\begin{equation}
P(\sigma \hat{\rho})=\frac{P(\sigma\hat{\rho} \mid n)}{\sum_{\hat{\rho}, \sigma} P(\sigma\hat{\rho} \mid n)},
\end{equation}
where $P(\sigma\hat{\rho} \mid n)$ is the probability of being at position $n$, proposing to move in the direction $\sigma\hat{\rho}$ and accepting the proposal (as described in steps 2 and 3 of the update step). Since $\sum_k P(\sigma\hat{\rho})=1$, the tail always moves. If there is a counter associated with the position $n$, then its value should be incremented by $\nicefrac{1}{\left(\sum_{\sigma\hat{\rho}} P(\sigma\hat{\rho} \mid n)\right)}$, the expected time, in the original algorithm, the tail is at $n$ before moving. In essence, we are replacing a stochastic step in the algorithm with its exact solution, which reduces the statistical errors. This update can also be understood as the heat bath step, from which the detailed balance follows. The comparison with the previous algorithm, shown in fig. \ref{fig:comparisonCT_vs_M}, demonstrates that the continuous time update represents a significant improvement over the standard algorithm.

\subsection{Ratio between correlation functions}\label{sec:ratio between correlation functions.}
First, let us show how operator insertions can be interpreted as background charges. Consider the correlation function $\left\langle e^{i Q \theta(x)} e^{-i Q \theta(y)}\right\rangle$. Then, by expanding the definition of the expectation value  
\begin{equation}
\left\langle e^{i Q \theta(x)} e^{-i Q \theta(y)}\right\rangle = \frac{\int\left(\prod_k d \theta_k\right) \exp \left[\beta \sum_{\langle i, j\rangle} \cos \left[\theta_i-\theta_j\right] + i Q\theta(x) - i Q\theta(y) \right]}{\int\left(\prod_k d \theta_k\right) \exp \left[\beta \sum_{\langle i, j\rangle} \cos \left[\theta_i-\theta_j\right]\right]},
\end{equation}
and using the identity 
\begin{equation}
\exp\left\{ \beta\cos\left(\theta_{i}-\theta_{j}\right)\right\} =\sum_{k_{ij}\in \mathbb{Z}}I_{k_{ij}}\left(\beta\right)e^{ik_{ij}\left(\theta_{i}-\theta_{j}\right)},
\end{equation}
the path integral over $\theta$ can be performed analytically, yielding 
\begin{equation}
\left\langle e^{i Q \theta(x)} e^{-i Q \theta(y)}\right\rangle = \dfrac{\sum_{\left\langle i,j\right\rangle}\sum_{ k_{ij} }I_{k_{ij}}\left(\beta\right)\delta\left(\sum_{i}\left(D_{i}+Q\delta_{ix}-Q\delta_{iy}\right)\right)}{\sum_{\left\langle i,j\right\rangle }\sum_{k_{ij}}I_{k_{ij}}\left(\beta\right)\delta\left(\sum_{i}\left(D_{i}\right)\right)},
\end{equation}
where $D_{i}\equiv\sum_{\hat{\rho} \in {\hat{1},\hat{2},\hat{3}}} k_{i,i+a\hat{\rho}} -k_{i-a\hat{\rho},i}$ and $\left\langle i,j\right\rangle$ is the sum over first neighbors.

We are now ready to study ratios between correlation functions. Consider the ratio appearing in eq.~\eqref{eq:decomposition 2pt function in ratios} 
\begin{align}\nonumber
\dfrac{\left\langle e^{iQ\theta(x)}e^{-iQ\theta(y)}\right\rangle }{\left\langle e^{i(Q-1)\theta(x)}e^{-i(Q-1)\theta(y)}\right\rangle } & =\dfrac{\sum_{k_{ij}}I_{k_{ij}}(\beta)\delta\left(\sum_{i}\left(D_{i}+\left(Q-1\right)\delta_{ix}-\left(Q-1\right)\delta_{iy}+\delta_{ix}-\delta_{iy}\right)\right)}{\sum_{k_{ij}}I_{k_{ij}}(\beta)\delta\left(\sum_{i}\left(D_{i}+\left(Q-1\right)\delta_{ix}-\left(Q-1\right)\delta_{iy}\right)\right)}\\
 & =\left\langle e^{i\theta\left(x\right)}e^{-i\theta\left(y\right)}\right\rangle _{\left(Q-1\right)_{x}-\left(Q-1\right)_{y}}
\end{align}

To implement this on the lattice, it suffices to generate an initial configuration of link variables satisfying 
\begin{equation}
\sum_i\left(D_i+(Q-1) \delta_{i x}-(Q-1) \delta_{i y}\right)
\end{equation}
at every point and then perform the standard worm update.  

This can be generalized to higher point functions. The general rule is that charge insertions in the denominator are removed from the numerator and added to the background. Thus, the 3pt function in eq.~\eqref{eq:continuum infinite size predictions} can be rewritten as
\begin{align}\nonumber
    \left\langle\mathcal{O}_{-Q}(x) \mathcal{O}_{2 Q}(0) \mathcal{O}_{-Q}(-x)\right\rangle&=\dfrac{\left\langle\mathcal{O}_{-Q}(x) \mathcal{O}_{2 Q}(0) \mathcal{O}_{-Q}(-x)\right\rangle}{\left\langle\mathcal{O}_Q(0) \mathcal{O}_{-Q}(-x)\right\rangle}\left\langle\mathcal{O}_Q(0) \mathcal{O}_{-Q}(-x)\right\rangle\\
    &=
\left\langle\mathcal{O}_{-Q}(x) \mathcal{O}_Q(0)\right\rangle_{Q_0-Q_{-x}}\left\langle\mathcal{O}_Q(0) \mathcal{O}_{-Q}(-x)\right\rangle
\end{align}

\section{Finite size scaling analysis}\label{sec:extrapolation}

In this appendix, we study the 
dependence of the numerical measurements of the conformal dimension on the size of the system. The data presented in the table below fig.\ref{fig:conformal dimension} are obtained using the procedure described here.

\begin{figure}[t]
    \centering
    \includegraphics[width=\columnwidth]{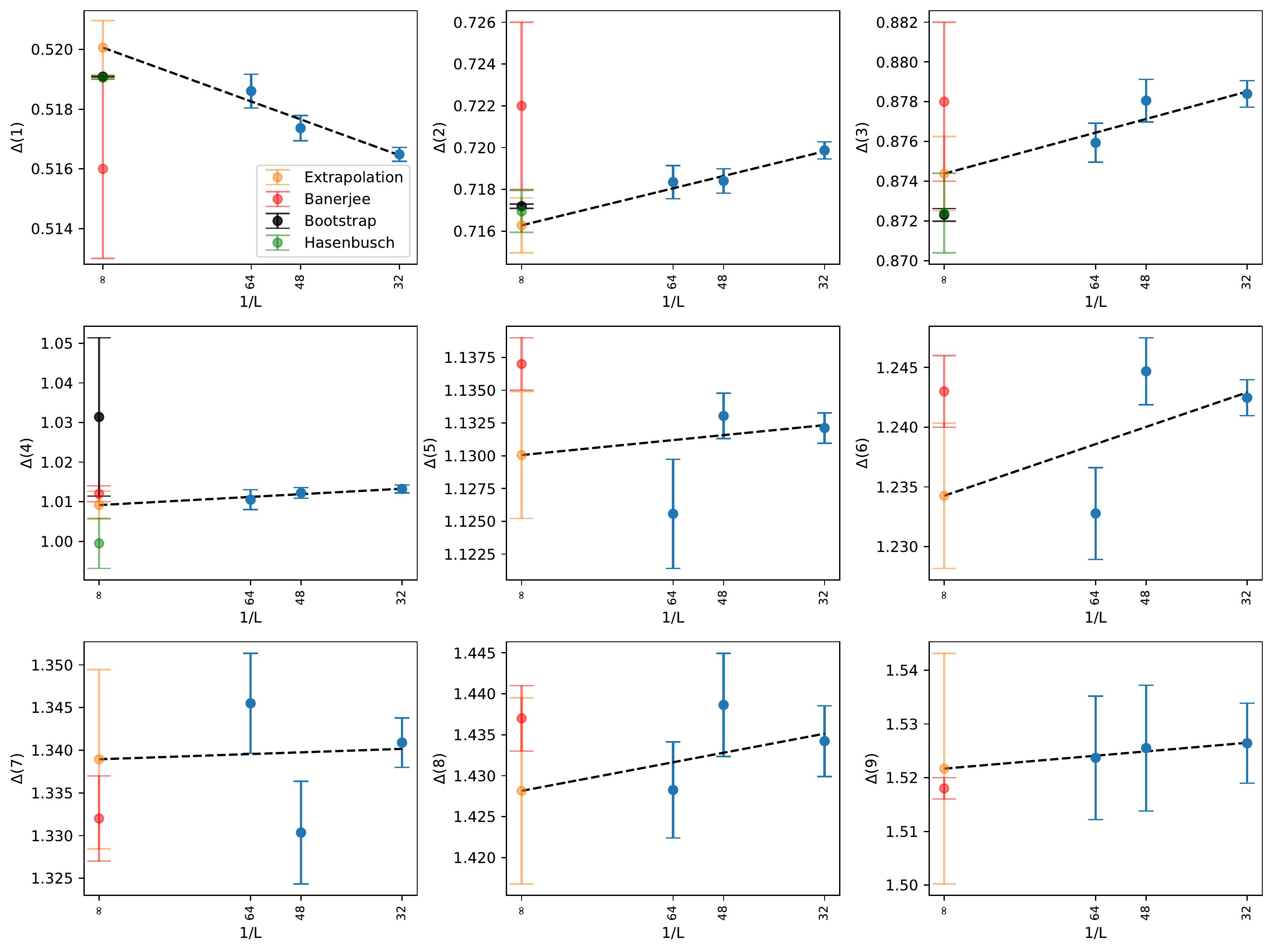}
    \caption{Extrapolations of the difference between consecutive conformal dimensions to $L=\infty$. In blue we present our data for $L=32,48,64$. In orange we present the linear extrapolation to $L=\infty$ and in red the results in \cite{Banerjee:2017fcx}. In black, we present the bootstrap results \cite{Chester:2019ifh}. In green, we present the results from \cite{PhysRevB.100.224517}, for $Q=1$, and \cite{PhysRevB.84.125136} for $Q=2,3,4$. Not all of the results exist for all the charges.}
    \label{fig:finite-size scaling}
\end{figure}

In fig. \ref{fig:finite-size scaling}, we show the numerical results for the differences $\Delta(Q)$ between conformal dimensions for $L=32,48$ and $64$ and their extrapolation to $L=\infty$, and present the comparison with the available results in the literature.
For $Q=1,2,3,4$, finite-size effects are relevant and bigger than statistical uncertainties. In particular, the measurement for  $Q=1$ and $L=32$ is incompatible with the bootstrap result \cite{Chester:2019ifh} and with previous Monte Carlo results \cite{PhysRevB.100.224517,PhysRevB.84.125136}, while our extrapolated value is compatible.\footnote{We thank Martin H. Hasenbusch for pointing out the mismatch with the bootstrap results in a previous version of this preprint.}

As charges become larger, systematic errors become less relevant and, for $Q=7$, the results for $L=32$ are compatible with the extrapolated results, within the uncertainties of the latter. 

In light of the above analysis, we made the following choice for the data presented in tab. \ref{tab: table conformal dimensions}: for charges $Q\leq7$, we use the extrapolated data and uncertainties; for $Q>7$ we use the data for $L=32$ with doubled uncertainties. We made this choice because, for $L=64$ and charge $Q\gtrsim10$, there are systematic errors due to the small statistics. Such systematic errors prevent us from obtaining reliable extrapolations to $L=\infty$.
From fig. \ref{fig:finite-size scaling} we observe that the statistical errors are roughly comparable with the systematic ones for $Q\geq7$, therefore justifying the choice of doubling the statistical uncertainties to estimate the overall uncertainty.

\section{Lattice corrections}\label{sec:lattice corrections}
The analysis we perform in sec.~\ref{sec:OPE coefficients} relies on the hypothesis that both lattice and finite size effects are small. However, we observe significant lattice effects for large charges, see fig.~\ref{fig:Ratio 3pt functions in the torus}, and thus linear extrapolation is no longer enough. Given that we are already using the largest lattice size and distances that we can reasonably simulate, we now study
lattice and finite size effects. 

 We start by identifying the projection of a lattice operator into continuum operators. Next, we identify neutral scalar irrelevant operators that can be added to the action. These need to be irrelevant otherwise the system flows away from this fixed point. Their goal is to encode the information about the existence of a lattice. Finally, our space has the topology of a torus, such that the 2pt functions are not fully fixed by symmetry. Thus, we only have access to correlation functions when operators are close, $|x - y|\ll L$, and the OPE quickly converges. Since this is related to the UV behaviour of theory, it is not sensitive to the global topology of the system.

\begin{table}[t]
    \centering
   \begin{tabular}{|c|c|c|c|}
\hline Name & Q & s & $\Delta$ \\
\hline \hline s & 0 & 0 & 1.511 \\
\hline $s^{\prime}$ & 0 & 0 & 3.789 \\
\hline $\phi$ & 1 & 0 & 0.519 \\
\hline $\phi^{\prime}$  & 1 & 0 & $\sim 4$ \\
\hline $t$ & 2 & 0 & 1.2361 \\
\hline $t^{\prime}$ & 2 & 0 & 3.624 \\
\hline $u$ & 3 & 0 & 2.1 \\
\hline $v$ & 4 & 0 & 3.1 \\
\hline $x$ & 5 & 0 & 4.26 \\
\hline
\end{tabular}
    \caption{Known conformal dimension of scalar operators in the 3d O(2) model \cite{Chester:2019ifh}.}
    \label{tab: table conformal dimensions}
\end{table}

For the sake of simplicity, we will only present explicit computations for the 2pt functions with $Q=1$. Generalizations to other charges or higher-order correlation functions are straightforward. We use the standard CFT notation. To make a connection with the rest of the paper, the reader should keep in mind that $\mathcal{O}_{\text {lat}}(x)=e^{i\theta (x)}$, $\Delta_\phi = D(1)$.

 The lightest charged scalar lattice operator can overlap with all operators that have the same charge and are scalar under the cubic subgroup (i.e. operators whose spin $s\equiv 0 \mod  4$). We will just consider the two lightest operators $\phi$ and $\square  \phi$, check Tab.\ref{tab: table conformal dimensions}. The lattice operator 
 \begin{equation}
 \mathcal{O}_{\text {lat}}=c_1 a^{\Delta_\phi} \phi+c_2 a^{\Delta_\phi+2} \square \phi,
 \end{equation}
 such that the 2pt function becomes 
 \begin{equation}\label{eq:lattice 2pt function}
 \left\langle\mathcal{O}^{\dagger}_\text{lat}(x) \mathcal{O}_{\text{lat}}(0)\right\rangle=a^{2 \Delta_\phi}\left(|c_1|^2  +2\text{Re}(c_2c_1) a^{2} \square_x\right)\left\langle\phi^{\dagger}(x) \phi(0)\right\rangle + \mathcal{O}\left(a^{3 + 2 \Delta_\phi}\right).
 \end{equation}

The next step is to deform the action. The lightest neutral scalar irrelevant operator is $s^\prime$ such that the first term in the deformed action is 
\begin{equation}
S=S_{\mathrm{CFT}}+g_{s^{\prime}} a^{\Delta_{s^{\prime}}-3} \int_{T^3} d^3 z s^{\prime}(z)+\cdots,
\end{equation}
where $g_{s^\prime}$ is a dimensionless parameter. By the usual perturbative expansion, the perturbed correlation function at one-loop is given by
\begin{equation}\label{eq:corrected 2pt function}
\left\langle\phi^{\dagger}(x) \phi(0)\right\rangle=\left\langle\phi^{\dagger}(x) \phi(0)\right\rangle_{\mathrm{CFT}}-g_{s^{\prime}} a^{\Delta_{s^{\prime}}-3} \int_{T^3} d^3 z\left\langle\phi^{\dagger}(x) \phi(0) s^{\prime}(z)\right\rangle_{\mathrm{CFT}}.
\end{equation}
The correlation functions on the right-hand side are computed on the unperturbed CFT.

In flat space, the 2pt and 3pt functions are known. In the torus, they are not and the only tool available is the OPE. This means we can only study the short-range behaviour of these correlation functions. In the OPE of $\phi^\dagger \times \phi$ we will only include the lightest neutral scalar operator \footnote{Descendants can also be considered. We do not include them here to keep the expressions manageable.}
\begin{equation}
\phi^{\dagger}(x) \times \phi(0) \sim|x|^{-2 \Delta_\phi}\left(\mathbb{I}+|x|^{\Delta_s} \mathcal{\lambda}_{\phi^{\dagger} \phi s}  s(0)\right).   
\end{equation}

Plugging this into \eqref{eq:corrected 2pt function}, we obtain 
\begin{align}\nonumber
\left\langle \phi^{\dagger}\left(x\right)\phi\left(0\right)\right\rangle =\left|x\right|^{-2\Delta_{\phi}} & \left(\left\langle \mathbb{I}\right\rangle _{\text{CFT}}+\left|x\right|^{\Delta_{s}}\lambda_{\phi^{\dagger}\phi s}\left\langle s\left(0\right)\right\rangle _{\text{CFT}}\right.\\
 & \left.-g_{s^{\prime}}a^{\Delta_{s^{\prime}}-3}\int_{T^{3}}d^{3}z\left[\left\langle s^{\prime}\left(z\right)\right\rangle _{\text{CFT}}+\left|x\right|^{\Delta_{s}}\lambda_{\phi^{\dagger}\phi s}\left\langle s\left(0\right)s^{\prime}\left(z\right)\right\rangle _{\text{CFT}}\right]\right),
\end{align}
which depends on vacuum expectation value (VEV) of $s$ and $s^\prime$ and in the integrated 2pt-function on the torus of $s$ and $s^\prime$. These are unknown in general,
hence, instead of focusing on computing them explicitly, we extract their dependence on the dimensionful parameter $L$. Let us go case by case:
\begin{itemize}
    \item $\left\langle \mathbb{I}\right\rangle_{\text{CFT}} = 1$, by definition of the identity operator.
    \item $\left\langle s\left(0\right)\right\rangle _{\text{CFT}} = \dfrac{\alpha_1}{L^{\Delta_s}}$, where $\alpha_1$ is a dimensionless parameter.
    $L$ appears raised to the power of the conformal dimension of $s$ since it is the only dimensionful parameter available\footnote{In $\mathbb{R}^3$ there is no such length scale, resulting in 1pt functions that are zero.}. 
    \item $\int_{T^{3}}d^{3}z\left\langle s^{\prime}\left(z\right)\right\rangle _{\text{CFT}} = \left\langle s^{\prime}\left(0\right)\right\rangle _{\text{CFT}} \int_{T^{3}}d^{3}z=\dfrac{\alpha_2}{L^{\Delta_{s^\prime} - 3}}$, where we used translation invariance to bring the expectation value of $s^\prime$ out of the integral.
    \item $\int_{T^{3}}d^{3}z\left\langle s\left(0\right)s^{\prime}\left(z\right)\right\rangle _{\text{CFT}} = \dfrac{\alpha_3}{L^{\Delta_s + \Delta_{s^\prime} - 3}}$ for the same reasons as before. 
\end{itemize}
Thus, we obtain the following perturbed 2pt-function
\begin{equation}
    \left\langle \phi^{\dagger}\left(x\right)\phi\left(0\right)\right\rangle =\left|x\right|^{-2\Delta_{\phi}}\left[1+\tilde{\alpha}_{1}\left(\dfrac{a}{L}\right)^{\Delta_{s^{\prime}}-3}+\tilde{\alpha}_{2}\left(\dfrac{\left|x\right|}{L}\right)^{\Delta_{s}}\left(1+\tilde{\alpha}_{3}\left(\dfrac{a}{L}\right)^{\Delta_{s^{\prime}}-3}\right)\right].
\end{equation}
By acting with $\square$ on this, we obtain the second term in eq. \eqref{eq:lattice 2pt function}.

Corrections to the OPE coefficients are obtained using the same ideas, but the derivations are significantly more cumbersome. As such, we only show the end result. Thus, the lattice estimation of the OPE coefficient appearing on the right-hand side of eq. \eqref{eq:OPE coefficients large charge function}, here denoted as $\lambda_{\text{OPE}}^{\left(\text{lat}\right)}$, is related with the "true" OPE coefficient, $\lambda_{\text{OPE}}$, as 
\begin{align}\nonumber
    \lambda_{\text{OPE}}^{\left(\text{lat}\right)} = \lambda_{\text{OPE}}
    &\left[1+\beta_{1}\left(\frac{a}{L}\right)^{\Delta_{s^{\prime}}-3}+\beta_2\left(\dfrac{x}{L}\right)^{\Delta_{s}}\left(1+\beta_{3}\left(\frac{a}{L}\right)^{\Delta_{s^{\prime}}-3}\right)\right.\\
    &\;\;\;\;\left.+\beta_{4}\left(\dfrac{a}{x}\right)^{2}\left(1+\beta_{5}\left(\dfrac{x}{L}\right)^{\Delta_{s}}\right)+\dots \right], \label{eq:lambda_OPE_lat}
\end{align}
where we kept all terms up to order $\left(\frac{a}{x}\right)^2$, $\left(\frac{a}{L}\right)^{\Delta_{s^{\prime}}-3}$ and $\left(\frac{x}{L}\right)^{\Delta_s}$, excluding mixed terms. This relation depends on the charges appearing on the left-hand side of eq. \eqref{eq:OPE coefficients large charge function} through OPE coefficients of the type $\lambda_{Q, -Q,s}$ and multiplicative factors of $\Delta_Q$ (these will never appear on the exponents of $x$ or $a$). The expansion~\eqref{eq:lambda_OPE_lat} is independent of the charges of the operators, up to the undetermined coefficients.

\bibliographystyle{JHEP}
\bibliography{main}

\providecommand{\href}[2]{#2}\begingroup\raggedright\begin{thebibliography}{10}

\bibitem{Rychkov:2016iqz}
S.~Rychkov, \emph{{EPFL Lectures on Conformal Field Theory in D\ensuremath{>}=
  3 Dimensions}}, SpringerBriefs in Physics (1, 2016),
  \href{https://doi.org/10.1007/978-3-319-43626-5}{10.1007/978-3-319-43626-5},
  [\href{https://arxiv.org/abs/1601.05000}{{\ttfamily 1601.05000}}].

\bibitem{Simmons-Duffin:2016gjk}
D.~Simmons-Duffin, \emph{{The Conformal Bootstrap}},  in \emph{{Theoretical
  Advanced Study Institute in Elementary Particle Physics}: {New Frontiers in
  Fields and Strings}}, pp.~1--74, 2017,
  \href{https://doi.org/10.1142/9789813149441_0001}{DOI}
  [\href{https://arxiv.org/abs/1602.07982}{{\ttfamily 1602.07982}}].

\bibitem{Alday:2007mf}
L.F.~Alday and J.M.~Maldacena, \emph{{Comments on operators with large spin}},
  \href{https://doi.org/10.1088/1126-6708/2007/11/019}{\emph{JHEP} {\bfseries
  11} (2007) 019} [\href{https://arxiv.org/abs/0708.0672}{{\ttfamily
  0708.0672}}].

\bibitem{Fitzpatrick:2012yx}
A.L.~Fitzpatrick, J.~Kaplan, D.~Poland and D.~Simmons-Duffin, \emph{{The
  Analytic Bootstrap and AdS Superhorizon Locality}},
  \href{https://doi.org/10.1007/JHEP12(2013)004}{\emph{JHEP} {\bfseries 12}
  (2013) 004} [\href{https://arxiv.org/abs/1212.3616}{{\ttfamily 1212.3616}}].

\bibitem{Komargodski:2012ek}
Z.~Komargodski and A.~Zhiboedov, \emph{{Convexity and Liberation at Large
  Spin}}, \href{https://doi.org/10.1007/JHEP11(2013)140}{\emph{JHEP} {\bfseries
  11} (2013) 140} [\href{https://arxiv.org/abs/1212.4103}{{\ttfamily
  1212.4103}}].

\bibitem{Hellerman:2015nra}
S.~Hellerman, D.~Orlando, S.~Reffert and M.~Watanabe, \emph{{On the CFT
  Operator Spectrum at Large Global Charge}},
  \href{https://doi.org/10.1007/JHEP12(2015)071}{\emph{JHEP} {\bfseries 12}
  (2015) 071} [\href{https://arxiv.org/abs/1505.01537}{{\ttfamily
  1505.01537}}].

\bibitem{Monin:2016jmo}
A.~Monin, D.~Pirtskhalava, R.~Rattazzi and F.K.~Seibold, \emph{{Semiclassics,
  Goldstone Bosons and CFT data}},
  \href{https://doi.org/10.1007/JHEP06(2017)011}{\emph{JHEP} {\bfseries 06}
  (2017) 011} [\href{https://arxiv.org/abs/1611.02912}{{\ttfamily
  1611.02912}}].

\bibitem{Caron-Huot:2017vep}
S.~Caron-Huot, \emph{{Analyticity in Spin in Conformal Theories}},
  \href{https://doi.org/10.1007/JHEP09(2017)078}{\emph{JHEP} {\bfseries 09}
  (2017) 078} [\href{https://arxiv.org/abs/1703.00278}{{\ttfamily
  1703.00278}}].

\bibitem{Simmons-Duffin:2017nub}
D.~Simmons-Duffin, D.~Stanford and E.~Witten, \emph{{A spacetime derivation of
  the Lorentzian OPE inversion formula}},
  \href{https://doi.org/10.1007/JHEP07(2018)085}{\emph{JHEP} {\bfseries 07}
  (2018) 085} [\href{https://arxiv.org/abs/1711.03816}{{\ttfamily
  1711.03816}}].

\bibitem{Hellerman:2017veg}
S.~Hellerman, S.~Maeda and M.~Watanabe, \emph{{Operator Dimensions from
  Moduli}}, \href{https://doi.org/10.1007/JHEP10(2017)089}{\emph{JHEP}
  {\bfseries 10} (2017) 089}
  [\href{https://arxiv.org/abs/1706.05743}{{\ttfamily 1706.05743}}].

\bibitem{Hellerman:2018xpi}
S.~Hellerman, S.~Maeda, D.~Orlando, S.~Reffert and M.~Watanabe,
  \emph{{Universal correlation functions in rank 1 SCFTs}},
  \href{https://doi.org/10.1007/JHEP12(2019)047}{\emph{JHEP} {\bfseries 12}
  (2019) 047} [\href{https://arxiv.org/abs/1804.01535}{{\ttfamily
  1804.01535}}].

\bibitem{Grassi:2019txd}
A.~Grassi, Z.~Komargodski and L.~Tizzano, \emph{{Extremal correlators and
  random matrix theory}},
  \href{https://doi.org/10.1007/JHEP04(2021)214}{\emph{JHEP} {\bfseries 04}
  (2021) 214} [\href{https://arxiv.org/abs/1908.10306}{{\ttfamily
  1908.10306}}].

\bibitem{Sharon:2020mjs}
A.~Sharon and M.~Watanabe, \emph{{Transition of Large $R$-Charge Operators on a
  Conformal Manifold}},
  \href{https://doi.org/10.1007/JHEP01(2021)068}{\emph{JHEP} {\bfseries 01}
  (2021) 068} [\href{https://arxiv.org/abs/2008.01106}{{\ttfamily
  2008.01106}}].

\bibitem{Komargodski:2021zzy}
Z.~Komargodski, M.~Mezei, S.~Pal and A.~Raviv-Moshe, \emph{{Spontaneously
  broken boosts in CFTs}},
  \href{https://doi.org/10.1007/JHEP09(2021)064}{\emph{JHEP} {\bfseries 09}
  (2021) 064} [\href{https://arxiv.org/abs/2102.12583}{{\ttfamily
  2102.12583}}].

\bibitem{Dondi:2022zna}
N.~Dondi, S.~Hellerman, I.~Kalogerakis, R.~Moser, D.~Orlando and S.~Reffert,
  \emph{{Fermionic CFTs at large charge and large N}},
  \href{https://arxiv.org/abs/2211.15318}{{\ttfamily 2211.15318}}.

\bibitem{Jafferis:2017zna}
D.~Jafferis, B.~Mukhametzhanov and A.~Zhiboedov, \emph{{Conformal Bootstrap At
  Large Charge}}, \href{https://doi.org/10.1007/JHEP05(2018)043}{\emph{JHEP}
  {\bfseries 05} (2018) 043}
  [\href{https://arxiv.org/abs/1710.11161}{{\ttfamily 1710.11161}}].

\bibitem{Zhu:2022gjc}
W.~Zhu, C.~Han, E.~Huffman, J.S.~Hofmann and Y.-C.~He, \emph{{Uncovering
  conformal symmetry in the $3D$ Ising transition: State-operator
  correspondence from a fuzzy sphere regularization}},
  \href{https://arxiv.org/abs/2210.13482}{{\ttfamily 2210.13482}}.

\bibitem{Loukas:2018zjh}
O.~Loukas, D.~Orlando, S.~Reffert and D.~Sarkar, \emph{{An AdS/EFT
  correspondence at large charge}},
  \href{https://doi.org/10.1016/j.nuclphysb.2018.07.020}{\emph{Nucl. Phys. B}
  {\bfseries 934} (2018) 437}
  [\href{https://arxiv.org/abs/1804.04151}{{\ttfamily 1804.04151}}].

\bibitem{Monin:2016bwf}
A.~Monin, \emph{{Partition function on spheres: How to use zeta function
  regularization}},
  \href{https://doi.org/10.1103/PhysRevD.94.085013}{\emph{Phys. Rev. D}
  {\bfseries 94} (2016) 085013}
  [\href{https://arxiv.org/abs/1607.06493}{{\ttfamily 1607.06493}}].

\bibitem{Cuomo:2020rgt}
G.~Cuomo, \emph{{A note on the large charge expansion in 4d CFT}},
  \href{https://doi.org/10.1016/j.physletb.2020.136014}{\emph{Phys. Lett. B}
  {\bfseries 812} (2021) 136014}
  [\href{https://arxiv.org/abs/2010.00407}{{\ttfamily 2010.00407}}].

\bibitem{Cuomo:2021ygt}
G.~Cuomo, \emph{{OPE meets semiclassics}},
  \href{https://doi.org/10.1103/PhysRevD.103.085005}{\emph{Phys. Rev. D}
  {\bfseries 103} (2021) 085005}
  [\href{https://arxiv.org/abs/2103.01331}{{\ttfamily 2103.01331}}].

\bibitem{Dondi:2022wli}
N.~Dondi, I.~Kalogerakis, R.~Moser, D.~Orlando and S.~Reffert, \emph{{Spinning
  correlators in large-charge CFTs}},
  \href{https://doi.org/10.1016/j.nuclphysb.2022.115928}{\emph{Nucl. Phys. B}
  {\bfseries 983} (2022) 115928}
  [\href{https://arxiv.org/abs/2203.12624}{{\ttfamily 2203.12624}}].

\bibitem{DeLaFuente:2018uee}
A.~De~La~Fuente, \emph{{The large charge expansion at large $N$}},
  \href{https://doi.org/10.1007/JHEP08(2018)041}{\emph{JHEP} {\bfseries 08}
  (2018) 041} [\href{https://arxiv.org/abs/1805.00501}{{\ttfamily
  1805.00501}}].

\bibitem{Badel:2019khk}
G.~Badel, G.~Cuomo, A.~Monin and R.~Rattazzi, \emph{{Feynman diagrams and the
  large charge expansion in $3-\varepsilon$ dimensions}},
  \href{https://doi.org/10.1016/j.physletb.2020.135202}{\emph{Phys. Lett. B}
  {\bfseries 802} (2020) 135202}
  [\href{https://arxiv.org/abs/1911.08505}{{\ttfamily 1911.08505}}].

\bibitem{Antipin:2022naw}
O.~Antipin, J.~Bersini and P.~Panopoulos, \emph{{Yukawa interactions at large
  charge}}, \href{https://doi.org/10.1007/JHEP10(2022)183}{\emph{JHEP}
  {\bfseries 10} (2022) 183}
  [\href{https://arxiv.org/abs/2208.05839}{{\ttfamily 2208.05839}}].

\bibitem{Badel:2019oxl}
G.~Badel, G.~Cuomo, A.~Monin and R.~Rattazzi, \emph{{The Epsilon Expansion
  Meets Semiclassics}},
  \href{https://doi.org/10.1007/JHEP11(2019)110}{\emph{JHEP} {\bfseries 11}
  (2019) 110} [\href{https://arxiv.org/abs/1909.01269}{{\ttfamily
  1909.01269}}].

\bibitem{Antipin:2020abu}
O.~Antipin, J.~Bersini, F.~Sannino, Z.-W.~Wang and C.~Zhang, \emph{{Charging
  the $O(N)$ model}},
  \href{https://doi.org/10.1103/PhysRevD.102.045011}{\emph{Phys. Rev. D}
  {\bfseries 102} (2020) 045011}
  [\href{https://arxiv.org/abs/2003.13121}{{\ttfamily 2003.13121}}].

\bibitem{Alvarez-Gaume:2019biu}
L.~Alvarez-Gaume, D.~Orlando and S.~Reffert, \emph{{Large charge at large N}},
  \href{https://doi.org/10.1007/JHEP12(2019)142}{\emph{JHEP} {\bfseries 12}
  (2019) 142} [\href{https://arxiv.org/abs/1909.02571}{{\ttfamily
  1909.02571}}].

\bibitem{Giombi:2020enj}
S.~Giombi and J.~Hyman, \emph{{On the large charge sector in the critical O(N)
  model at large N}},
  \href{https://doi.org/10.1007/JHEP09(2021)184}{\emph{JHEP} {\bfseries 09}
  (2021) 184} [\href{https://arxiv.org/abs/2011.11622}{{\ttfamily
  2011.11622}}].

\bibitem{Banerjee:2017fcx}
D.~Banerjee, S.~Chandrasekharan and D.~Orlando, \emph{{Conformal dimensions via
  large charge expansion}},
  \href{https://doi.org/10.1103/PhysRevLett.120.061603}{\emph{Phys. Rev. Lett.}
  {\bfseries 120} (2018) 061603}
  [\href{https://arxiv.org/abs/1707.00711}{{\ttfamily 1707.00711}}].

\bibitem{Prokofev:2001ddj}
N.~Prokof'ev and B.~Svistunov, \emph{{Worm Algorithms for Classical Statistical
  Models}}, \href{https://doi.org/10.1103/PhysRevLett.87.160601}{\emph{Phys.
  Rev. Lett.} {\bfseries 87} (2001) 160601}
  [\href{https://arxiv.org/abs/cond-mat/0103146}{{\ttfamily
  cond-mat/0103146}}].

\bibitem{Banerjee:2010kc}
D.~Banerjee and S.~Chandrasekharan, \emph{{Finite size effects in the presence
  of a chemical potential: A study in the classical non-linear O(2)
  sigma-model}}, \href{https://doi.org/10.1103/PhysRevD.81.125007}{\emph{Phys.
  Rev. D} {\bfseries 81} (2010) 125007}
  [\href{https://arxiv.org/abs/1001.3648}{{\ttfamily 1001.3648}}].

\bibitem{Banerjee:2019jpw}
D.~Banerjee, S.~Chandrasekharan, D.~Orlando and S.~Reffert, \emph{{Conformal
  dimensions in the large charge sectors at the O(4) Wilson-Fisher fixed
  point}}, \href{https://doi.org/10.1103/PhysRevLett.123.051603}{\emph{Phys.
  Rev. Lett.} {\bfseries 123} (2019) 051603}
  [\href{https://arxiv.org/abs/1902.09542}{{\ttfamily 1902.09542}}].

\bibitem{Banerjee:2021bbw}
D.~Banerjee and S.~Chandrasekharan, \emph{{Subleading conformal dimensions at
  the O(4) Wilson-Fisher fixed point}},
  \href{https://doi.org/10.1103/PhysRevD.105.L031507}{\emph{Phys. Rev. D}
  {\bfseries 105} (2022) L031507}
  [\href{https://arxiv.org/abs/2111.01202}{{\ttfamily 2111.01202}}].

\bibitem{Singh:2022akp}
H.~Singh, \emph{{Large-charge conformal dimensions at the $O(N)$ Wilson-Fisher
  fixed point}},  \href{https://arxiv.org/abs/2203.00059}{{\ttfamily
  2203.00059}}.

\bibitem{Kos:2015mba}
F.~Kos, D.~Poland, D.~Simmons-Duffin and A.~Vichi, \emph{{Bootstrapping the
  O(N) Archipelago}},
  \href{https://doi.org/10.1007/JHEP11(2015)106}{\emph{JHEP} {\bfseries 11}
  (2015) 106} [\href{https://arxiv.org/abs/1504.07997}{{\ttfamily
  1504.07997}}].

\bibitem{Dondi:2021buw}
N.~Dondi, I.~Kalogerakis, D.~Orlando and S.~Reffert, \emph{{Resurgence of the
  large-charge expansion}},
  \href{https://doi.org/10.1007/JHEP05(2021)035}{\emph{JHEP} {\bfseries 05}
  (2021) 035} [\href{https://arxiv.org/abs/2102.12488}{{\ttfamily
  2102.12488}}].

\bibitem{Hellerman:2021duh}
S.~Hellerman, \emph{{On the exponentially small corrections to ${\cal N} = 2$
  superconformal correlators at large R-charge}},
  \href{https://arxiv.org/abs/2103.09312}{{\ttfamily 2103.09312}}.

\bibitem{Cuomo:2017vzg}
G.~Cuomo, A.~de~la Fuente, A.~Monin, D.~Pirtskhalava and R.~Rattazzi,
  \emph{{Rotating superfluids and spinning charged operators in conformal field
  theory}}, \href{https://doi.org/10.1103/PhysRevD.97.045012}{\emph{Phys. Rev.
  D} {\bfseries 97} (2018) 045012}
  [\href{https://arxiv.org/abs/1711.02108}{{\ttfamily 1711.02108}}].

\bibitem{Cuomo:2022kio}
G.~Cuomo and Z.~Komargodski, \emph{{Giant Vortices and the Regge Limit}},
  \href{https://doi.org/10.1007/JHEP01(2023)006}{\emph{JHEP} {\bfseries 01}
  (2023) 006} [\href{https://arxiv.org/abs/2210.15694}{{\ttfamily
  2210.15694}}].

\bibitem{Ning_informal}
N.~Su and J.~Rong, ``{Private Communication}.''

\bibitem{Gribov:2003nw}
V.N.~Gribov, \emph{{The theory of complex angular momenta: Gribov lectures on
  theoretical physics}}, Cambridge Monographs on Mathematical Physics,
  Cambridge University Press (6, 2007),
  \href{https://doi.org/10.1017/CBO9780511534959}{10.1017/CBO9780511534959}.

\bibitem{Simmons-Duffin:2016wlq}
D.~Simmons-Duffin, \emph{{The Lightcone Bootstrap and the Spectrum of the 3d
  Ising CFT}}, \href{https://doi.org/10.1007/JHEP03(2017)086}{\emph{JHEP}
  {\bfseries 03} (2017) 086}
  [\href{https://arxiv.org/abs/1612.08471}{{\ttfamily 1612.08471}}].

\bibitem{Liu:2020tpf}
J.~Liu, D.~Meltzer, D.~Poland and D.~Simmons-Duffin, \emph{{The Lorentzian
  inversion formula and the spectrum of the 3d O(2) CFT}},
  \href{https://doi.org/10.1007/JHEP09(2020)115}{\emph{JHEP} {\bfseries 09}
  (2020) 115} [\href{https://arxiv.org/abs/2007.07914}{{\ttfamily
  2007.07914}}].

\bibitem{Correia:2020xtr}
M.~Correia, A.~Sever and A.~Zhiboedov, \emph{{An analytical toolkit for the
  S-matrix bootstrap}},
  \href{https://doi.org/10.1007/JHEP03(2021)013}{\emph{JHEP} {\bfseries 03}
  (2021) 013} [\href{https://arxiv.org/abs/2006.08221}{{\ttfamily
  2006.08221}}].

\bibitem{Caron-Huot:2020ouj}
S.~Caron-Huot, Y.~Gobeil and Z.~Zahraee, \emph{{The leading trajectory in the
  2+1D Ising CFT}}, \href{https://doi.org/10.1007/JHEP02(2023)190}{\emph{JHEP}
  {\bfseries 02} (2023) 190}
  [\href{https://arxiv.org/abs/2007.11647}{{\ttfamily 2007.11647}}].

\bibitem{ballesterosFiniteSizeEffects1996}
H.G.~Ballesteros, L.A.~Fernandez, V.~Martin-Mayor and A.~Munoz~Sudupe,
  \emph{{Finite size effects on measures of critical exponents in d = 3 O(N)
  models}}, \href{https://doi.org/10.1016/0370-2693(96)00984-7}{\emph{Phys.
  Lett. B} {\bfseries 387} (1996) 125}
  [\href{https://arxiv.org/abs/cond-mat/9606203}{{\ttfamily
  cond-mat/9606203}}].

\bibitem{Prokof_ev_2001}
N.~Prokof{\textquotesingle}ev and B.~Svistunov, \emph{Worm algorithms for
  classical statistical models},
  \href{https://doi.org/10.1103/physrevlett.87.160601}{\emph{Physical Review
  Letters} {\bfseries 87} (2001) }.

\bibitem{Chester:2019ifh}
S.M.~Chester, W.~Landry, J.~Liu, D.~Poland, D.~Simmons-Duffin, N.~Su et~al.,
  \emph{{Carving out OPE space and precise $O(2)$ model critical exponents}},
  \href{https://doi.org/10.1007/JHEP06(2020)142}{\emph{JHEP} {\bfseries 06}
  (2020) 142} [\href{https://arxiv.org/abs/1912.03324}{{\ttfamily
  1912.03324}}].

\bibitem{PhysRevB.100.224517}
M.~Hasenbusch, \emph{Monte carlo study of an improved clock model in three
  dimensions}, \href{https://doi.org/10.1103/PhysRevB.100.224517}{\emph{Phys.
  Rev. B} {\bfseries 100} (2019) 224517}.

\bibitem{PhysRevB.84.125136}
M.~Hasenbusch and E.~Vicari, \emph{Anisotropic perturbations in
  three-dimensional o($n$)-symmetric vector models},
  \href{https://doi.org/10.1103/PhysRevB.84.125136}{\emph{Phys. Rev. B}
  {\bfseries 84} (2011) 125136}.

\end{thebibliography}\endgroup

\end{document}